\documentclass{aa}  
\usepackage{graphicx}
\usepackage{txfonts}
\usepackage{natbib}
\usepackage{hyperref}
\hypersetup{colorlinks=True,citecolor=blue,allcolors=blue}
\usepackage{tabularx}
\usepackage{tabu}
\usepackage{float}
\pdfoutput=1 

\begin{document}

   \title{Continuum reverberation mapping of the quasar PG 2130+099}

\author{C. Fian\inst{1}, D. Chelouche\inst{2}, S. Kaspi\inst{1}, C. Sobrino Figaredo\inst{3}, S. Catalan\inst{2}, T. Lewis\inst{4}}%
\institute{School of Physics and Astronomy and Wise Observatory, Raymond and Beverly Sackler Faculty of Exact Sciences, Tel-Aviv University, Tel-Aviv, Israel \and Haifa Research Center for Theoretical Physics and Astrophysics, University of Haifa, Haifa, Israel \and Astronomisches Institut Ruhr-Universit\"at Bochum, Universit\"atsstraße 150, D-44801 Bochum, Germany \and NASA Postdoctoral Program Fellow, NASA Goddard Space Flight Center, Code 661, 8800 Greenbelt Rd, Greenbelt, MD 20771, USA}
 

  \abstract
  {}
   {We present the results of an intensive six-month optical continuum reverberation mapping campaign  of the Seyfert 1 galaxy PG 2130+099 at redshift $z=0.063$. The ground-based photometric monitoring was conducted on a daily basis with the robotic 46 cm telescope of the Wise observatory located in Israel. Specially designed narrowband filters were used to observe the central engine of the active galactic nucleus (AGN), avoiding line  contamination from the broad-line region (BLR). We aim to measure inter-band continuum time lags across the optical range and determine the size-wavelength relation for this system.}
   {We used two methods, the traditional point-spread function (PSF) photometry and the recently developed proper image subtraction technique, to independently perform the extraction of the continuum light curves. The inter-band time lags are measured with several methods, including the interpolated cross-correlation function, the z-transformed discrete correlation function, a von Neumann estimator, JAVELIN (in spectroscopic mode), and MICA.}
   {PG 2130+099 displays correlated variability across the optical range, and we successfully detect significant time lags of up to $\sim3$ days between the multiband light curves. We find that the wavelength-dependent lags, $\tau(\lambda)$, generally follow the relation $\tau(\lambda) \propto \lambda^{4/3}$, as expected for the temperature radial profile $T \propto R^{-3/4}$ of an optically thick, geometrically thin accretion disk. Despite that, the derived time lags can also be fitted by $\tau(\lambda) \propto \lambda^2$, implying the possibility of a slim, rather than thin, accretion disk. 
   Using the flux variation gradient (FVG) method, we determined the AGN's host-galaxy-subtracted rest frame 5100\AA\ luminosity at the time of our monitoring campaign with an uncertainty of $\sim$18\% ($\lambda L_{5100} = (2.40\pm0.42) \times 10^{44}$ erg s$^{-1}$).
   While a continuum reprocessing model can fit the data reasonably well, our derived disk sizes are a factor of $\sim2-6$ larger than the theoretical disk sizes predicted from the AGN luminosity estimate of PG 2130+099. This result is in agreement with previous studies of AGN/quasars and suggests that the standard Shakura-Sunyaev disk theory has limitations in describing AGN accretion disks.}
   {}


\titlerunning{Continuum reverberation mapping of PG 2130+099}
\authorrunning{Fian et al.} 
\maketitle

\section{Introduction}
Active galactic nuclei (AGNs) are astrophysical sources powered by the accretion of material onto a galaxy's central super-massive black hole (SMBH). The material around a SMBH orbits in a plane around the center, forming a so-called accretion disk that produces multi-temperature black-body emission (in the classical Shakura-Sunyaev picture), peaking typically in the rest-ultraviolet (rest-UV) wavelength range. In current standard models, the accretion disk is considered to be geometrically thin and optically thick (\citealt{Shakura1973}). Thus, the disk will be hotter at inner radii and cooler toward outer radii (\citealt{Shields1978}), following a temperature gradient of $T(R) \propto R^{-3/4}$ over a large range of $R$, where $R$ is the distance from the central SMBH (assuming that the disk luminosity is produced by black-body radiation). In this scenario, the hotter, UV-emitting region is expected to be located closer to the center, while the cooler, optically emitting regions are located farther out. For such a temperature profile (since $\tau \sim R/c$ and $\lambda \propto 1/T$; from Wien's law), the disk sizes are wavelength dependent and scale as $R_{\lambda} \propto \lambda^{\ \beta}$. The exponent $\beta$ equals 4/3 in the standard thin disk model, whereas for a slim disk ($T(R) \propto R^{-1/2}$; \citealt{Wang1999}) the wavelength-dependent sizes follow $R_{\lambda} \propto \lambda^2$ in the inner parts of the disk. \\

Understanding the size and structure of AGNs is crucial for enhancing our knowledge about the physics of their accretion disks (\citealt{Dexter2019,Wilkins2020}), their kinematics, and their geometry (\citealt{Motta2012,Motta2017,Fian2018a,Guerras2013a,Guerras2013b,Popovic2020,Rojas2020}), for estimating the mass of their central SMBHs (\citealt{Assef2011,Mediavilla2018,Mediavilla2019,Mediavilla2020,Homayouni2020}) and studying their growth and evolution (\citealt{Goulding2010,Peng2006}), and for obtaining better insight into the connection between various emission components (\citealt{Lobban2020}). One difficulty that arises when studying AGNs is that we are currently unable to directly resolve their innermost regions because they are spatially too small. However, we can use alternative methods to indirectly infer the structure of different (continuum) emitting regions and probing temperature profiles using either microlensing in lensed AGNs (\citealt{Kochanek2004,JimenezVicente2012,JimenezVicente2014,Fian2016,Fian2018b,Fian2021a,Fian2021b,Fian2021c}) or reverberation mapping (RM; \citealt{PozoNunez2017,PozoNunez2019,Chan2020,Yu2020}).\\

RM is a proven technique for determining the time-delay response between the variability of the ionizing continuum of an AGN and the line emission associated with the broad-line region (BLR; \citealt{Blandford1982}). Since the time delay is directly related to the size of the BLR, RM can be used for studying the structure and kinematics of BLRs in AGNs. A further use of RM - due to the technique's independence of spatial resolution - is mapping the continuum emitting region around the SMBH at the center of an AGN (\citealt{Wanders1997,Collier1998}). The basic idea of continuum RM is to measure the time lag, $\tau$, between continuum bands at different wavelengths. Assuming that the variation in the continuum emission across the continuum source(s) is triggered by the variation of a central illuminating source (such as in the lamp-post reprocessing picture; see, e.g., \citealt{Cackett2007}), the variation at longer wavelengths is expected to lag the variation at shorter wavelengths due to the light-crossing time between the hotter, inner disk and the cooler, outer disk. The light-travel time between the two respective emission sites then is 
\begin{equation}
\centering
\tau = \frac{R_{\lambda_0}}{c}\left[\left(\frac{\lambda}{\lambda_0}\right)^\beta-1\right],
\label{eq1}
\end{equation}
where $R_{\lambda_0}$ is the effective disk size at wavelength $\lambda_0$.
Measuring the wavelength dependence of those time lags in correlated continuum light curves therefore gives both the size scale and its temperature profile for a previously assumed disk geometry.\\

In this work, we focus on the Seyfert 1 galaxy PG 2130+099 ($m_V = 14.3$, $z = 0.063$; see \citealt{Grier2013}) and present continuum emitting region size measurements using photometric monitoring data obtained at the Centurion 18 inch telescope (C18) located at the Wise observatory. The paper is organized as follows. In Section \ref{2} we describe the observations and detail the data reduction. We outline the methodology and time series analysis in Section \ref{3}, where we compare several tools to measure the time delays between light curves taken in multiple photometry bands. The results and discussion are presented in Section \ref{4}, including the continuum time lags, the lag spectrum, the host-subtracted AGN luminosity, the theoretical disk size, transfer functions, and directionality measurements. Finally, a conclusion is given in Section \ref{6}.

\section{Observations and data reduction}\label{2}
\begin{figure}[h]
\centering
\includegraphics[width=9cm]{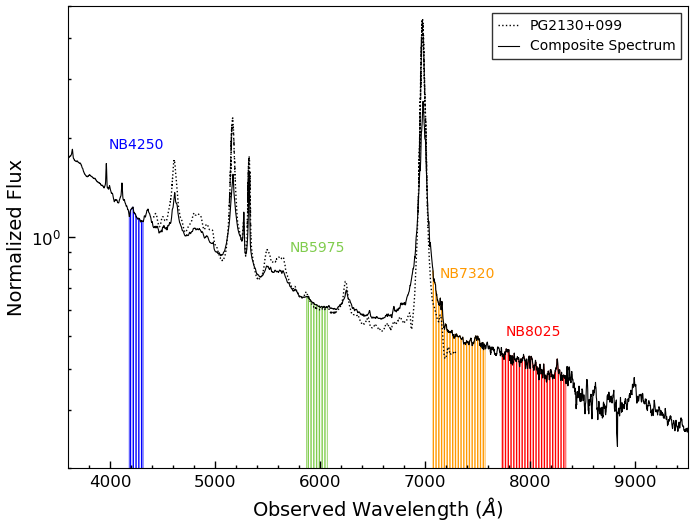}
\caption{Composite AGN spectrum of \citealt{Glikman2006} (solid line; shown at the quasar's redshift at $z = 0.063$), overlaid with the spectrum of PG 2130+099 obtained by \citealt{Grier2008} (dotted line). The four emission-line-free NB passes of the filters which mainly trace the AGN continuum variations and are used for the photometric monitoring presented in this work, are shaded in different colors (blue for NB4250, green for NB5975, orange for NB7320, and red for NB8025).}
\label{spectrum}
\end{figure}

The photometric monitoring was carried out between June 2019 and December 2019 at the Wise Observatory located near the town Mitzpe-Ramon in Israel. The observations were conducted on a daily basis ($\sim$4 exposures per night in each filter) for a duration of almost six months using the robotic C18 telescope (\citealt{Brosch2008}). We used the QSI 683 CCD (image sensor KAF-8300), which has $3326\times2504$ pixels of size $5.4\mu m$. The pixel scale is 0.882 arcsec\,pix$^{-1}$, which gives a field of view of $48.9\times36.8$ arcmin ($0.815\times0.613$ degrees, which corresponds to an area of 0.5 deg$^2$). In Table \ref{characteristics} we list the object's characteristics, and in Table \ref{obs_fil} we summarize the filter and observation information. The range of airmass covered during each night of the object's observations was between 1.1 and 2.0. To trace the AGN continuum variations free of emission-lines at the quasar's restframe, four relatively narrow bands (NBs) located at 4250, 5975, 7320, and 8025\AA\ were carefully chosen. Figure \ref{spectrum} shows the position of the NB filters together with the spectrum of PG 2130+099 obtained by \citet{Grier2008}, superimposed on a quasars' composite spectrum (\citealt{Glikman2006}). The images were reduced following standard procedures performed with IRAF, including bias subtraction, dark current correction, and flat fielding for each filter. We used two independent methods (PSF photometry and proper image subtraction) to obtain the light curves as outlined below. 
\begin{table}[h]
	\tabcolsep=0cm
	\renewcommand{\arraystretch}{1.2}
	\caption{PG 2130+099 characteristics.}
	\begin{tabu} to 0.493\textwidth {X[c]X[c]X[c]X[c]X[c]} 

		\hline
		\hline 
		R.A. (J2000.0) & Dec. (J2000.0) & $m_V$ & $z$ & $\lambda L_\lambda$ (5100\AA)\\  
		(1) & (2) & (3) & (4) & (5)\\ \hline
		$21\ 32\ 27.8$ & $+10\ 08\ 19$ & $14.64$ & $0.063$ & $2.40\pm0.42$ \\ \hline 
	\end{tabu} \\

		\small NOTES. --- Cols. (1)--(2): Right ascension and declination from NED. Units of right ascension are hours, minutes, and seconds, and units of declination are degrees, arcminutes, and arcseconds. Col. (3): $V$ magnitude from the \citealt{Veron-Cetty2010} catalog. Col. (4): Redshift from \citet{Grier2013}. Col. (5): Host-subtracted AGN luminosity interpolated to restframe 5100\AA\ (in units of $10^{44}$ ergs s$^{-1}$; see Section \ref{host} for a detailed discussion).
\label{characteristics}	
\end{table}

\begin{table*}[h]
	\tabcolsep=0cm
	\renewcommand{\arraystretch}{1.2}
	\caption{Filter and observation information.}
	\begin{tabu} to \textwidth {X[c]X[c]X[c]X[c]X[c]X[c]} 

		\hline
		\hline 
		Filter & CWL (\AA) & $\lambda$ (\AA) & $\Delta \lambda$ (\AA) & $t_{exp}$ (s) & S/N\\  
		(1) & (2) & (3) & (4) & (5) & (6)\\ \hline
		NB4250 & 4250 & $4150-4330$ & 138 & $400$ & $\sim10$ \\
		NB5975 & 5975 & $5850-6090$ & 188 & $280$ & $\sim6$\\
		NB7320 & 7320 & $7040-7610$ & 488 & $240$ & $\sim6$\\
		NB8025 & 8025 & $7700-8370$ & 588 & $220$ & $\sim5$ \\\hline 
	\end{tabu}\\ 

		\small NOTES. --- Col. (1): NB filter. Cols. (2)--(4): central wavelength, minimum and maximum wavelength at 0.01\% transmission, and FWHM in the units of \AA. Col. (5): Exposure time in seconds. Col. (6): typical signal-to-noise of PG 2130+099 combining four exposures.
\label{obs_fil}	
\end{table*}

\subsection{PSF photometry}
The first method used in this work is the traditional point-spread function (PSF) fitting photometry. We used the DAOPHOT (\citealt{Stetson1987}) package as implemented in IRAF and DAOSTAT (\citealt{Netzer1996}) to measure the objects in the images and to compute the light curve of the quasar, respectively. 
First,  DAOPHOT is used to define an empirical PSF for all objects and to measure the instrumental magnitude for each object including the quasar, in each image. Then,
the  objects in each image are aligned in order to have the same (x,y) coordinates according to a predefined reference image.
In the second stage DAOSTAT is used to create the light curve. 
This program first checks each potential star for real variability or for bad data points owing to bad pixels, cosmic rays, and other features. This is done by taking one of the images as a reference image and choosing one of the stars as a reference star. The chosen star should be (i) non-variable, (ii) appear in all the other frames, and (iii) not be too bright to avoid saturation problems. 
Then, for each star in each image the mean magnitude relative to the chosen reference star is computed. Once this is done for all stars an improved mean is calculated by comparing with the mean magnitude of all the other stars in the same image. 
For each star DAOSTAT constructs the light curve and computes the mean value and its standard deviation, $\sigma$. 
This allows us to identify stars that show variations or have bad points and they can be removed from the data set. 
We now have a set of comparison stars and can remove variable stars or stars that do not have very accurate measurements, according to the $\sigma$ of the light curve.
The removal of stars is repeated until we are left with a high-quality set of $\sim$15-30 comparison stars with $\sigma\leq0.01$. The larger the number of images and number of comparison stars in the field of the target, the better is the accuracy achieved. Finally, we can define a zero point for the magnitude-scale of each image including error estimations. Using the scaling of each image and the use of a weighted linear regression of instrumental versus scaled magnitudes of the standard stars, we scale also the instrumental magnitude of the quasar in each image to get its light curve. 

\subsection{Proper image subtraction}
\begin{figure}[h]
\centering
\includegraphics[width=8cm]{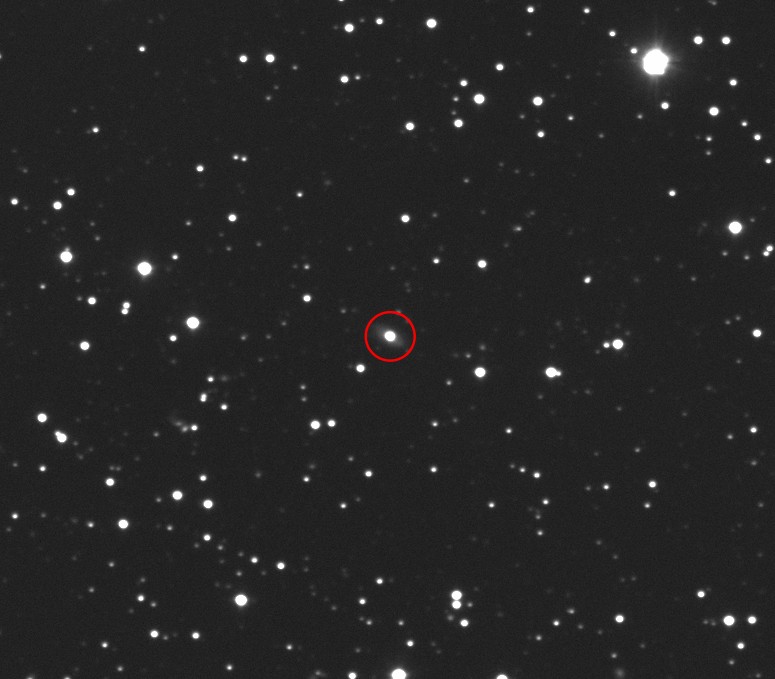}
\caption{Portion of the reference image ($11\times10$ arcmin) containing the Seyfert 1 galaxy PG 2130+099 built from the individual NB7320 frames collected throughout the monitoring program using the 46cm telescope of the Wise observatory in Israel. The images have been registered, interpolated, and stacked with \textit{SWarp}.}
\label{referenceimage}
\end{figure}

The second method used in this work to derive the quasar's light curve in each filter is the proper image subtraction (\citealt{Zackay2016}). One of the main advantages of image subtraction techniques is the removal of the extended host galaxy which in many cases contributes to the observed AGN flux. A sensible approach to remove the non-varying host and non-varying AGN contributions is to subtract a high-signal-to-noise reference image. Prior to the subtraction one has to accomplish the following steps:

\begin{itemize}
\item[1.] \textit{Registration and interpolation.}
We apply \textit{SWarp} (\citealt{Bertin1996,Bertin2010}) to accurately register positions of the sources with respect to a reference frame. During this process, each image frame is resampled and interpolated from the instrumental pixel size to a new common coordinate grid. The \textit{SWarp} registrations make use of \textit{SExtractor} to identify all the objects in the field, and afterward for the performance of centroiding algorithms.
\item[2.] \textit{Quality image selection.}
In order to create a set of good images, we have to sort out the images according to their quality. Images that contain condensation rings, elongated stars due to tracking or auto-guider problems, bright streaks across the image due to satellites, or other artifacts are rejected. Also images with poor transmission and exposures that are contaminated by scattered light due to the presence of the moon or light from passing cars were discarded. After carefully inspecting all the images, we are left with $\sim$70\% of the frames on average (depending on the filter; for more details see Table \ref{info}) to be coadded to produce the reference frame. \vspace*{1mm}
\item[3.] \textit{Background subtraction.}
The sky background in wide-field-of-view astronomical images cannot be treated as constant over the entire field of view. Since the background is a composition of instrumental noise and light diffusion from the atmosphere (from both incoming light and man-made light pollution), it is prone to changes between different exposures. If not subtracted, the stacking of all those exposures will produce a patchwork created by the different individual backgrounds. Since large-scale gradients are commonly found on astronomical images, subtraction of a constant from each frame will generally yield poor results. A solution to this problem is to subtract a smooth background map that contains low-spatial-frequency noise components of the data, including offsets. This can be done by using \textit{SWarp} routines which compute an estimator for the local background in each mesh of grid that covers the whole frame. A good compromise has to be found for the mesh size (in pixel), since the background estimation for too small meshes is affected by the presence of objects and random noise while too large mesh sizes will not be able to reproduce all the variations in the background. We used a mesh size of 128 pixels, large enough to dilute the flux of extended objects in the background map. In addition, median filtering was applied to suppress possible local overestimations due to bright stars or artifacts. The final background map computed by \textit{SWarp} for each frame is a bicubic-spline interpolation between the meshes of the grid and is automatically subtracted from the input image.\vspace*{1mm}
\item[4.] \textit{Building a reference image.}
Given a good set of registered, interpolated, background-subtracted images, we use \textit{SWarp} to stack the images and build a high-signal-to-noise reference image (see Figure \ref{referenceimage}).\\
\end{itemize}

To carry out the subtraction and to perform photometry on the subtracted frames, we used the proper image subtraction code developed by \citet{Zackay2016} and several custom written tools. In this method, the high-quality reference image is convolved with a (spatially variable) kernel to match the PSF of each individual frame. The image subtraction procedures as implemented in the code of \citet{Zackay2016} can be summarized as follows. The first step is the (local) estimation, interpolation and subtraction of the background for all the individual frames and the high S/N reference image. The code uses the Matlab function \textit{mextractor} to perform a careful background subtraction, to extract the sources in the field, and to estimate the PSF for each input image. The PSF is likely not constant spatially since it sensitively depends on the focus (which can vary across the camera), tracking of the telescope as well as atmospheric conditions during the exposure (which blurs the stellar images). The simplest approach is to divide the image to smaller pieces in which the PSF is approximately constant. Also, a precise alignment to get rid of the shift between images is a critical step for any image subtraction technique since any leftover registration imperfection residuals will lead to improper subtraction and subtraction artifacts. In a next step the relative, flux-based zero points between the reference image and the individual frames are estimated. The algorithm then convolves the reference frame with a spatially variable convolution kernel in order to match it to the PSF of the input images. In the end, the convolved reference frame is subtracted from the convolved individual images, resulting in difference images only containing the fluxes from variable sources while all the constant sources are removed. Positive fluxes represent an excess and negative fluxes a deficit with respect to the flux of the reference image. Prior to the subtraction we trimmed the images (from 3300$\times$2500 pixels$^2$ to 1900$\times$1400 pixels$^2$) since it is recommended to remove the image-edges and to perform the subtraction only on a small image patch to minimize residuals (\citealt{Zackay2016}). Using custom-written tools we obtained the accurate coordinates of the AGN and performed aperture photometry on the previously obtained series of difference images. We specified several apertures and by using the \textit{Von Neumann Mean Square Successive Difference} (\citealt{VonNeumann1941}) we quantified which aperture delivers the lowest white noise component to the nuclear fluxes. Using too small apertures yields larger scores since we are only considering a small portion of the total flux. On the other hand, the score for too large apertures increases owing to a bigger contamination by the residual background. We find that apertures between 5 and 8 pixels (corresponding to $\sim4.4$ to $7.1\arcsec$; larger apertures for redder filters) deliver the best result.

\begin{table}[h]
\renewcommand{\arraystretch}{1.2}
\caption{Information about the reference images.}
\begin{tabu} to 0.493\textwidth {X[c]X[c]X[c]X[c]} 

\hline
\hline 
Filter & $N_{obs}$ & $N_{ref}$ & Percentage \\
(1) & (2) & (3) & (4) \\
\hline 
NB4250 & 556 & 471 & 85\% \\ 
NB5975 & 542 & 356 & 66\% \\ 
NB7320 & 536 & 343 & 64\% \\
NB8025 & 531 & 354 & 67\% \\ \hline 
\end{tabu}\\
	
	\small NOTES. --- Col. (1): NB filter. Cols. (2)--(4): total number of observations, number of images used for building the reference image, and corresponding percentage.
\label{info}    
\end{table}
\subsubsection{Error estimation for the proper image subtraction}
We calculated the total error $\sigma_{tot}$ similar to the DAOPHOT's error estimation by combining a background-only error, $\sigma_{bkg}$ (Gaussian background error estimated from a priorly defined sky annulus) with Poisson noise of the source $I$ (depending on the aperture and the effective gain $g_{eff}$),
\begin{equation}
\sigma_{tot} = \sqrt{\sigma_{bkg}^2 + \frac{I}{g_{eff}}}.
\end{equation}

\subsection{Light curve comparison}
\begin{figure*}
\centering
\includegraphics[width=14.2cm]{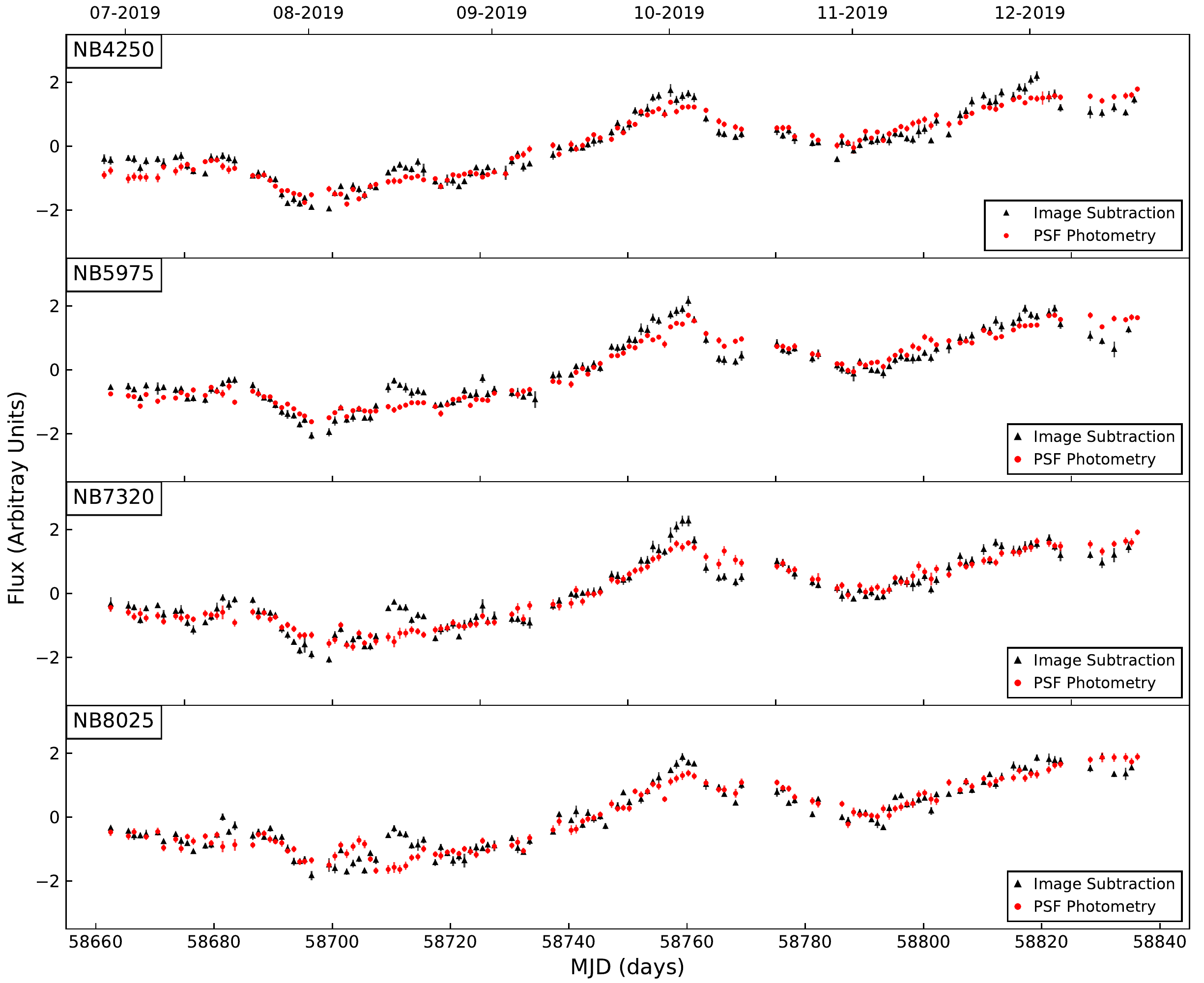}
\caption{From top to bottom: PSF photometry (red circles) and proper image subtraction (black triangles) light curves of the AGN continuum at 4250, 5975, 7320, and 8025\AA\ for the period between June 2019 and December 2019. The light curves are normalized to zero mean and unit standard deviation, and the fluxes are in arbitrary units.}
\label{lightcurves}
\end{figure*}

To obtain accurate measures for the flux at a given epoch for the two methods above, we discarded problematic exposures (due to low S/N, telescope tracking issues, or condensation rings on the CCD) from each night. To do so, we compared consecutive points and removed points above a certain threshold. The last step is to average the outlier-free exposures for each night. Finally, we were left with $\sim$130 data points (i.e., nights) for each light curve ($\sim$10 data points have been discarded). To directly compare the two methods, we show the normalized to mean and unit standard deviation light curves$^1$ \footnote{\hspace*{-3mm}$^1 $\ The light curves are available in electronic form
at the CDS via anonymous ftp to cdsarc.u-strasbg.fr (130.79.128.5)
or via http://cdsweb.u-strasbg.fr/cgi-bin/qcat?J/A+A/.}in Figure \ref{lightcurves} for the different bands. We can immediately see that the overall shapes of the light curves obtained with both methods are very similar. One possible explanation for the small differences between the methods is that the PSF photometry might enclose a contribution from the extended host galaxy while for the image subtraction the host galaxy has been successfully removed during the subtraction process and only the variable nucleus is left. \citet{Feng2017} found that for extended sources (as is the case here), the change of seeing may affect the aperture and therefore the contribution of the host. If the host galaxy is considered to be constant, the photometric results could reflect a combined contribution of this aperture effect and the intrinsic variation of the AGN itself. However, the comparison of the proper image subtraction method with the (host-contaminated) PSF photometry shows that the effects of seeing do not induce significant changes in the light curves. In Table \ref{var} we present the variability measures for all PSF photometry light curves. 
We note that the fractional variability amplitude ($F_{var}$) listed in Table \ref{var} is nearly constant in all bands while a smaller variability amplitude is expected at longer wavelengths. This can be explained by the increasing host contribution in the red filters (see Section \ref{host}).

\begin{table*}[h]
	\tabcolsep=0cm
	\renewcommand{\arraystretch}{1.2}
	\caption{Variability measures for the host-subtracted PSF photometry light curves.}
	\begin{tabu} to 1\textwidth {X[c]X[c]X[c]X[c]X[c]X[c]X[c]X[c]} 
		
		\hline
		\hline 
		Filter & $\overline{m}$ & $rms$ & $\delta$ & $\chi^2_\nu$ & $\sigma_N$ & $F_{var}$ & $Err\, F_{var}$ \\  
		 (1) & (2) & (3) &(4) & (5) & (6) & (7) & (8)\\ \hline
		NB4250 & $4.64$ & $0.44$ & $0.12$ & $17.0$ & $9.0$ & $0.090$ & $0.006$ \\
		NB5975 & $3.39$ & $0.34$ & $0.09$ & $16.2$ & $9.7$ & $0.097$ & $0.006$ \\ 
		NB7320 & $3.01$ & $0.32$ & $0.09$ & $15.2$ & $10.2$ & $0.102$ & $0.007$ \\ 
		NB8025 & $3.00$ & $0.31$ & $0.10$ & $9.6$ & $9.7$ & $0.097$ & $0.007$ \\ \hline 
	\end{tabu}\\
	
	\small NOTES. --- Col. (1): NB filter. Cols. (2)--(4): mean ($\overline{m}$), rms, and mean uncertainty ($\delta$) of all data points in the light curves in units of mJy. Col. (5): $\chi^2_\nu$ obtained by fitting a constant to the light curve. Col. (6): intrinsic normalized variability measure, $\sigma_N = 100\sqrt{(rms^2-\delta^2)}/\overline{m}$. Cols. (7)--(8): fractional variability amplitude and its uncertainty (\citealt{Rodriguez-Pascual1997,Edelson2002}).
\label{var}	
\end{table*}


\section{Time series analysis}\label{3}
The main goal of this work is to determine the continuum time lags between the NBs located at 4250, 5975, 7320, and 8025\AA. To robustly estimate the reverberation lags, we employed several methods that are commonly used for that purpose.

\begin{itemize}
\item[(a)] \textit{ICCF.}
The first method is the traditional interpolated cross-correlation function (ICCF) of \citet{Gaskell1986} and \citet{Gaskell1987}, as implemented by \citet{White1994}; see also review by \citet{Gaskell1994}. Due to the uneven sampling of the light curves, an interpolation is required to properly perform any cross-correlation function (CCF) analysis. The ICCF relies on a "two-way" linear interpolation over data gaps, and the CCF is calculated twice for the two observed light curves $a(t_i)$ and $b(t_i)$: once by pairing the observed $a(t_i)$ with the interpolated value $b(t_i-\tau)$, and once by pairing the observed $b(t_i)$ with the interpolated value $a(t_i-\tau)$. The average of those two CCFs provides the final ICCF. The time lag is then determined by measuring the centroid $\tau_{cent}$ of the points around the ICCF peak $r_{max}$, defined by
\begin{equation}
\tau_{cent} = \frac{\int \tau\ r(\tau)\ d\tau}{\int r(\tau)\ d\tau}
\end{equation}
for $r(\tau) \ge 0.6 r_{max}$ or $r(\tau)\ge 0.8 r_{max}$, where $r(\tau)$ is the correlation coefficient.\\

A major disadvantage of this method is the lack of a rigorous error estimation for the CCF and the inferred time lags (\citealt{Kaspi1996}). One way to estimate the errors is to use the flux randomization and random subset selection (FR/RSS) method of \citet{Peterson1998,Peterson2004}, described below in (c).\\

\item[(b)] \textit{ZDCF.}
A way to avoid interpolation is to use a discrete correlation function (DCF; \citealt{Edelson1988}), which evaluates the correlation function in bins of time delay. Here we used the z-transformed discrete correlation function (ZDCF) of \citealt{Alexander1997}, which is an improvement of the DCF. The ZDCF applies Fisher's z-transformation to the correlation coefficients, and uses equal population bins instead of equal time bins used in the DCF, yielding more robust and statistically reliable results. Again we take the centroid of the correlation function above 60\% or 80\% of the peak to measure the time lags between the light curves. The errors are estimated by a maximum likelihood method that takes into account the uncertainty in the ZDCF points.\\

\item[(c)] \textit{FR/RSS.}
To estimate the uncertainty in the ICCF time lags, we used the model-independent Monte Carlo flux randomization/random subset selection (FR/RSS) method (\citealt{Maoz1989,Peterson1998,Peterson2004}), a procedure that accounts for both the measurement errors of the fluxes and the uncertainties due to the sampling/cadence. For each light curve, 1000 sample light curves are produced by taking a random subset of the observed data points (RSS) and shifting the fluxes randomly (FR) according to their uncertainties by adding Gaussian noise. We then use ICCF to produce cross-correlation centroid distributions (CCCDs) for the Monte Carlo simulations. We allowed for a maximum time shift of $\pm$30 days and interpolated the light curves in 0.1 days steps. The median of the CCCD is taken as the final time lag and the width of the distribution is a measure of the uncertainty in the lag. The range that contains 68\% of the Monte Carlo realizations in the CCCD corresponds to 1$\sigma$ uncertainty for a normal distribution, meaning that the lower and upper errors of the lag are determined by the 16\% and 84\% quantiles of the CCCD.\\

\item[(d)] \textit{Von Neumann Estimator.}
\citet{Chelouche2017} recently introduced a method that does not depend on interpolation or binning (nor stochastic modeling) of the light curves but is based on the regularity of randomness of the data. The approach relies on the fact that quasar light curves carry information and attempts to find the relative time shift between two time series that minimizes the level of randomness. They found that the von Neumann's statistical estimator (\citealt{VonNeumann1941}) provides better lag measurements for irregularly sampled light curves where the underlying variability processes cannot be modeled properly. A more detailed explanation of the performance of this method is given in \citet{Chelouche2017} and is beyond the scope of the present paper.\\

\item[(e)] \textit{JAVELIN.}
Another popular and widely used method to measure reverberation lags is JAVELIN developed by \citet{Zu2011,Zu2013}. JAVELIN models the quasar's driving continuum light curve assuming a Damped Random Walk (DRW) process (\citealt{Kelly2009,MacLeod2010,MacLeod2012,Kozlowski2010,Kozlowski2016}) using two parameters, the amplitude and timescale of the quasar's stochastic variability. JAVELIN uses a Markov Chain Monte Carlo (MCMC) approach to maximize the likelihood of simultaneously fitting a DRW model to the light curves, assuming that the longer wavelength light curves are a shifted, scaled, and smoothed version of the driving light curve. JAVELIN then returns a lag posterior distribution from 10000 MCMC simulations, which is used to compute the lags and their uncertainties.\\

\item[(f)] \textit{MICA.}
Another algorithm used to measure lags is MICA (\citealt{Li2013,Li2016}), a forward-modeling method that also uses the DRW model to describe the variations in the driving continuum light curve and directly infers the transfer functions (by presuming a specific shape). While JAVELIN adopts a top-hat transfer function, MICA expresses the transfer function as a sum of a series of Gaussians. The time lags are given at the center of the Gaussians and the associated uncertainties are estimated as the standard deviation of the generated Markov chains.

\end{itemize}
\section{Results and discussion}\label{4}
In the following section, we present the time series analysis results for the Seyfert 1 galaxy PG 2130+099, including the inferred continuum time lags between the different NBs, the corresponding lag spectrum, an estimate for the host-subtracted AGN luminosity, the theoretical disk size as a function of luminosity, the recovered transfer functions, and an attempt to measure directionality in the quasar's light curves.

\subsection{Continuum time lags}
\begin{figure*}[!ht]
\centering
\includegraphics[width=15.4cm]{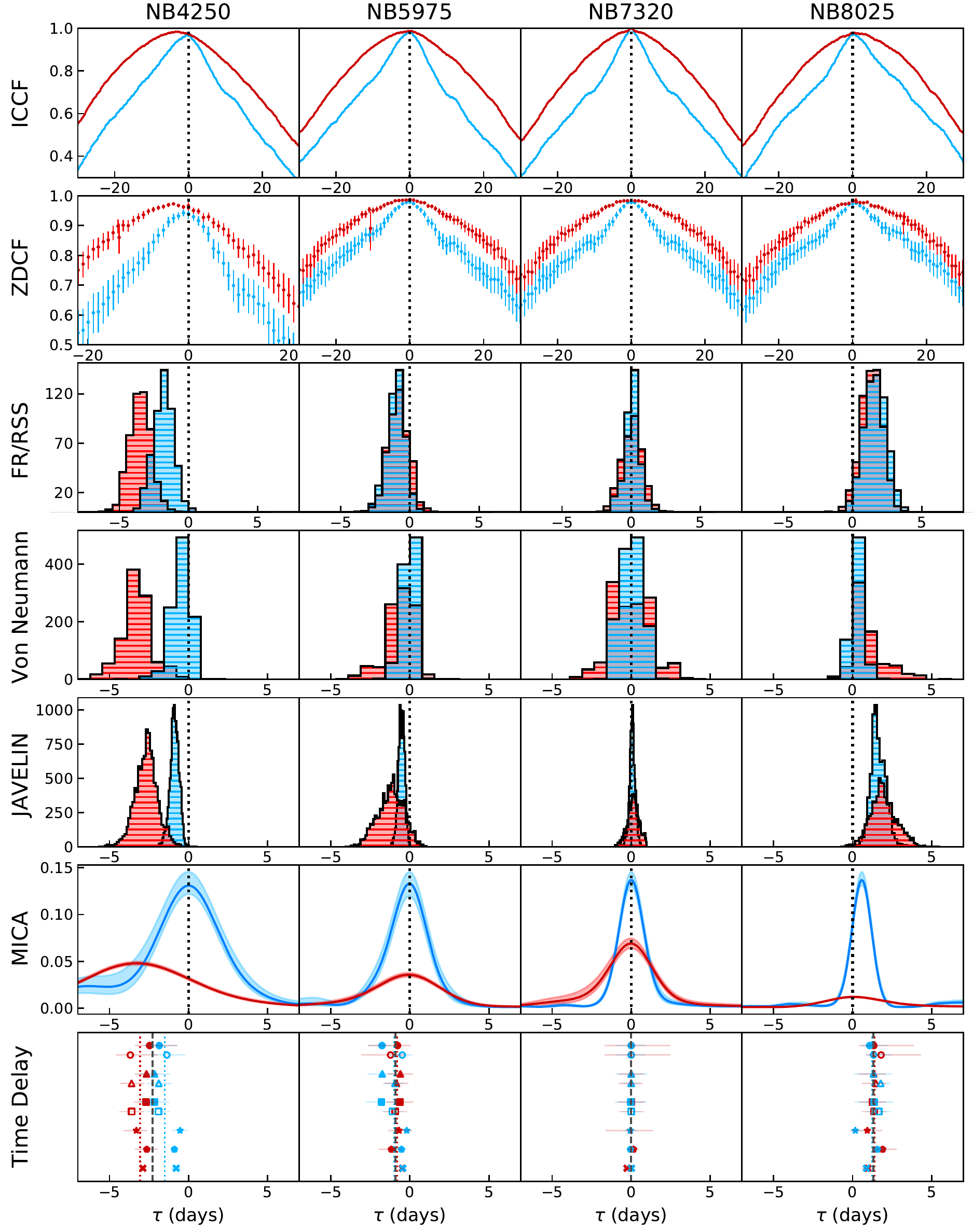}
\caption{From top to bottom: partially interpolated CCFs, z-transformed DCFs, FR/RSS centroid distributions (for centroid $\ge 0.8\ r_{max}$), von Neuman estimator peak distributions, JAVELIN posterior distributions of lags, and MICA transfer functions for each NB relative to the 7320\AA\ band. A positive value means the comparison band lags behind the NB7320 light curve. The blue distributions/functions stand for the results obtained by applying the methods to the proper image subtraction light curves while the red color corresponds to the results deduced from the PSF photometry light curves. In the bottom panel all time delays are plotted (in the same order as described before) on a vertical axis for illustration. The colored dotted lines show the mean time delay of each reduction procedure (i.e. PSF photometry and proper image subtraction) and the gray dashed line gives the mean time delay of all methods and reduction procedures together. These values are presented in Table \ref{list_td}.}
\label{timedelays}
\end{figure*}

In Figure \ref{timedelays}, we show the lag measurements for the four continuum light curves obtained for the various methods discussed in Section \ref{3}. All lags were measured with respect to the NB7320 light curve since this band shows the highest S/N and exhibits the least noisy shapes. This noise has been quantified by estimating the Neumann's Mean Square Successive Difference (\citealt{VonNeumann1941}), a measure of randomness over time. 
For validation purposes we also include the lag estimations of the 7320\AA\ NB relative to itself, resulting in symmetric distributions concentrated around zero as expected. From Figure \ref{spectrum} we see that the NB7320 filter might be affected to some extent by the red wing of H$\alpha$. However, our results are consistent when using the NB8025 filter as pivot instead, which is free of any broad-emission line wing. For each reduction method, the ICCF and ZDCF methods show consistent results and RM lags are clearly seen from the cross correlation analysis (except for the reference light curve). This indicates that the interpolation done in the ICCF did not introduce any artificial correlation. As can be seen for the JAVELIN posterior distributions and MICA transfer functions, also the light curve modeling techniques are able to capture reverberation lags, although leading to slightly smaller values for the proper image subtraction light curves. For both methods (JAVELIN and MICA) we used a common time-delay search interval $[\tau_{min},\tau_{max}] = [-10,10]$ days. The distribution of time delay obtained from the von Neumann's method after Monte Carlo simulation of FR/RSS as done for the ICCF analysis yields similar results to those derived from cross-correlation/modeling for the PSF photometry, but smaller values for the proper image subtraction.\\

\begin{table*}[!ht] 
\tabcolsep=0cm
\renewcommand{\arraystretch}{1.41}
\caption{Summary of the time lags expressed in light days in the observer's frame between the four continuum light curves of PG 2130+099.}
\begin{tabu} {X[c]X[c]X[c]X[c]X[c]} 

\hline
\hline 
Method & Filter$^a$ & PSF Photometry & Image Subtraction & Mean $\pm$ d$^b$\\ 
\hline \hline

ICCF ( $\ge0.8\ r_{max}$) & NB4250 &  $-3.6_{-0.7}^{+0.8}$ & $-1.9_{-0.7}^{+0.8}$ & $-2.7 \pm 0.9$\\ 
 & NB5975 &  $-0.9_{-0.8}^{+0.8}$ & $-0.9_{-0.8}^{+0.5}$ & $-0.9 \pm 0.1$\\ 
 & NB8025 &  $+1.5_{-0.9}^{+0.6}$ & $+1.8_{-0.6}^{+0.6}$ & $+1.6 \pm 0.2$\\ \hline


ZDCF ( $\ge0.8\ r_{max}$) & NB4250 & $-3.7_{-0.9}^{+1.7}$ & $-1.4_{-1.0}^{+1.2}$ & $-2.5 \pm 1.2$\\
 & NB5975 & $-1.2_{-1.9}^{+0.9}$ & $-0.5_{-0.9}^{+0.6}$ & $-0.9 \pm 0.4$\\
 & NB8025 & $+1.8_{-0.9}^{+2.5}$ & $+1.3_{-0.5}^{+1.1}$ & $+1.6 \pm 0.2$\\ \hline


FR/RSS ($\ge0.8 r_{max}$) & NB4250 & $-3.6_{-0.8}^{+0.8}$ & $-1.9_{-0.7}^{+0.7}$ & $-2.8 \pm 0.8$\\
 & NB5975 & $-0.9_{-0.8}^{+0.8}$ & $-1.1_{-0.7}^{+0.7}$ & $-1.0 \pm 0.1$\\
 & NB8025 & $+1.4_{-0.8}^{+0.8}$ & $+1.7_{-0.8}^{+0.8}$ & $+1.5 \pm 0.1$\\ \hline

Von Neumann Estimator & NB4250 & $-3.3_{-0.8}^{+0.7}$ & $-0.5_{-0.5}^{+0.5}$ & $-1.9 \pm 1.4$\\
 & NB5975 & $-0.7_{-0.7}^{+0.8}$ & $-0.2_{-0.2}^{+0.3}$ & $-0.4 \pm 0.2$\\
 & NB8025 & $+0.9_{-1.1}^{+0.9}$ & $+0.2_{-0.3}^{+0.5}$ & $+0.6 \pm 0.4$\\ \hline

JAVELIN & NB4250 & $-2.6_{-0.8}^{+0.7}$ & $-0.9_{-0.2}^{+0.2}$ & $-1.8 \pm 0.9$ \\
& NB5975 & $-1.2_{-0.8}^{+0.7}$ & $-0.5_{-0.3}^{+0.3}$ & $-0.9 \pm 0.3$\\
& NB8025 & $+1.9_{-0.7}^{+0.9}$ & $+1.6_{-0.4}^{+0.5}$ & $+1.7 \pm 0.2$\\ \hline

MICA & NB4250 & $-2.9_{-0.2}^{+0.2}$ & $-0.8_{-0.3}^{+0.3}$ & $-1.8 \pm 1.1$\\ 
 & NB5975 & $-0.5_{-0.2}^{+0.2}$ & $-0.4_{-0.2}^{+0.2}$ & $-0.5 \pm 0.1$\\ 
 & NB8025 & $+1.0_{-0.4}^{+0.4}$ & $+0.9_{-0.2}^{+0.2}$ & $+0.9 \pm 0.1$\\ \hline \hline

Mean All Methods$^{c}$ & NB4250 & $-3.1 \pm 0.5$ & $-1.5 \pm 0.6$ & $-2.3 \pm 0.8$\\ 
 & NB5975 & $-0.8 \pm 0.3$ & $-1.0 \pm 0.6$ & $-0.9 \pm 0.1$\\ 
& NB8025 & $+1.4 \pm 0.3$ & $+1.2 \pm 0.5$ & $+1.3 \pm 0.1$\\ \hline \hline 
\end{tabu}
\label{list_td} 
\vspace*{1mm}
$^a\ $relative to NB7320\\
$^b$ half the difference between PSF photometry and image subtraction\\
$^c$ to estimate the mean also values $\ge 0.6\ r_{max}$ have been used (not listed here) 
\end{table*}

Table \ref{list_td} summarizes the results for the PSF photometry and the proper image subtraction for each cross correlation method. For the ICCF, ZDCF, and FR/RSS we list the centroid time lag computed from all points within 80\% of the peak value $r_{max}$. We also estimated the time lags using all points above 60\% of the peak value, obtaining similar results. The right most column lists the mean time delay $\pm 1 \sigma$ of both reduction methods, and the three last rows contain the overall mean time delay of all time lag determination methods together. For the bluest filter (NB4250), there is a discrepancy between the PSF photometry and proper image subtraction, whereas for the other filters the deduced values for the time delay are in excellent agreement. Combining all the values listed in Table \ref{list_td}, we obtain the mean time delays in the observer's frame, $\tau = -2.3 \pm 0.6$ days for NB4250, $\tau = -0.9 \pm 0.4$ days for NB5975, and $\tau = +1.3 \pm 0.4$ days for NB8025, relative to the NB7320 light curve. The uncertainties have been obtained by estimating the standard deviation from the mean of all methods. Applying a weighted mean we obtain slightly smaller values for the time delays ($\tau=-2.0$ days for NB4250, $\tau=-0.5$ days for NB5975, and $\tau= +1.1$ days for NB8025), but consistent within errors with the time lags obtained when using the ordinary mean.\\

In the case of little short-timescale variability and significant variability on the timescales comparable to the duration of the observing campaign, aliasing becomes an increasing problem (\citealt{Peterson2014}). In the context of AGN variability studies, quasi-parabolic trends longer than reverberation timescales can yield misleading reverberation results. \citet{Welsh1999} found that time delays measured by cross-correlation are biased too low and subject to a large variance due to finite-duration sampling and the dominance of long timescale trends in the light curves, rather than noise or irregular sampling. Since the standard definition of the CCF implicitly assumes that the time series are stationary in their mean and variance, one can help to fulfill this requirement by removing the low-frequency power from the light curves prior to the ICCF analysis. \citet{Welsh1999} suggested that lag estimates can be substantially improved by "detrending" the light curves: fitting the light curves with low-order polynomials, and subtraction of these longer-term trends reduces the bias toward underestimating lags. However, detrending is necessary only if the light curves show evident secular variability, for example in the case of continuum and emission line light curves that exhibit different long-term trends (as the exact response of the line depends on the detailed structure and dynamics of the BLR; see \citealt{Peterson2002}). While for some systems removing the bias due to secular variability led to marked improvements in the results (\citealt{Denney2010,Li2013}), several authors (see, e.g., \citealt{Grier2008} in the particular case of PG 2310+099; \citealt{Hlabathe2020}) did not find any considerable improvements when applying detrending. In this paper, to test the impact of long-term secular variability (that is not associated with reverberation variations) on the lag estimates, we adopted a first-order polynomial to fit the PSF photometry and image subtraction light curves and then detrend the light curves by removing the polynomial. In all cases, we find that detrending has almost no effect on the ICCF results (with differences of $\Delta \tau \sim 0.1$ days on average). 

\subsection{Lag spectrum}
\begin{figure}[h]
\centering
\includegraphics[width=8.9cm]{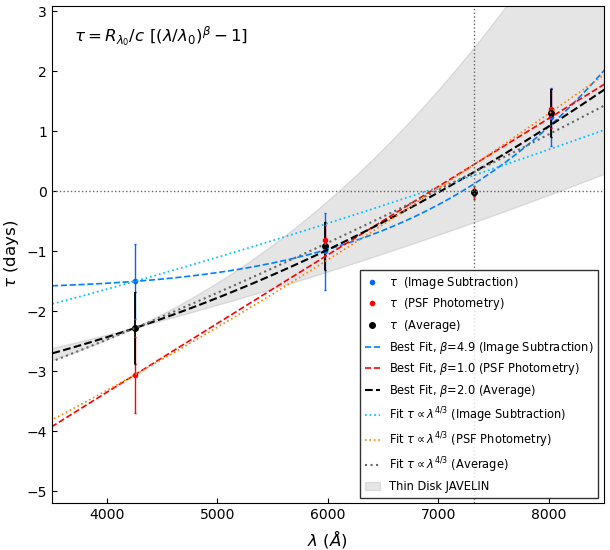}
\caption{PSF photometry (red), proper image subtraction (blue), and mean (black) time lags as a function of wavelength. All lags are measured with respect to the 7320\AA\ band. The dashed lines show the best fit to the observed relation $\tau= R_{\lambda_0}/c\ [\left(\lambda/\lambda_0\right)^{\ \beta}-1]$ with $R_{\lambda_0}/c$ and $\beta$ as free parameters. The dotted lines are fits with fixed theoretical power-law index $\beta = 4/3$, as expected for an optically thick and geometrically thin disk. The shaded envelope shows the results of the JAVELIN Thin Disk inspired model.}
\label{r_vs_lambda}
\end{figure}

Figure \ref{r_vs_lambda} shows the average time delay as a function of wavelength. Simply fitting the observed continuum lags in the four different photometric NBs with a disk model can provide the accretion disk size at a given wavelength. We first fitted our average delay spectrum to Equation \ref{eq1} with $R_{\lambda_0}/c$ and $\beta$ as free parameters. The best fit is obtained with $R_{\lambda_0}/c=1.3$ days and $\beta=2.0$. We then fixed the power-law index $\beta = 4/3$ in order to test the time delay-wavelength relation as predicted for an optically thick and geometrically thin accretion disk model, obtaining the best fit with $R_{\lambda_0}/c = 2.4$ days. We also show the results obtained from JAVELIN's thin disk inspired model, recently developed by \citet{Mudd2018}. This model fits directly for the disk parameters that best reproduces the given light curves of known wavelengths instead of providing a lag, top-hat width, and top-hat scale for each light curve. Both fits (and within uncertainties also the JAVELIN model) overlap and produce consistent results for the size of the disk, although the observed time lag at the reddest wavelength may favor a slightly steeper slope than predicted by the standard disk temperature profile. In previous RM studies several authors observed a steep rise in $\tau$ at long wavelengths (see, e.g., \citealt{Gaskell2007,Cackett2018,Chelouche2019}) and attributed this to contamination by light being reprocessed from further away. 

\subsection{Host-subtracted AGN luminosity}\label{host}
To determine the 5100\AA\ AGN luminosity, which is widely used in reverberation studies, the contribution of the host galaxy to the nuclear flux has to be subtracted. We disentangle the constant host from the variable AGN flux inside our aperture by using the flux variation gradient (FVG) method originally proposed by \citet{Choloniewski1981}, and further established by \citet{Winkler1992} and \citet{Sakata2010}. The absolute flux calibration is carried out on the stacked reference images by comparison with the PANSTARSS catalog. For each calibration star, we fit a black-body curve between the known $griz$ values and interpolate the flux to obtain the value for the NB filters. In addition, a Galactic foreground extinction correction (\citealt{Schlafly2011}) was applied. For multiband AGN monitoring data, \citet{Choloniewski1981} observed a linear relation between the fluxes obtained in two different filters taken at different epochs. He demonstrated the existence of a two-component structure to the spectra of Seyfert galaxies, with one component (the host) being constant in time and the second component (the AGN) showing strong variability. Despite the strong variability, the spectral energy distribution (SED) of the AGN does not change over time (\citealt{Pozo2012}). In the FVG method, data points for two filters obtained in the same night are plotted in a flux-flux diagram (in units of mJy). As the observed source varies in luminosity, the fluxes in the FVG diagram will follow a linear relation with a slope (denoted by the symbol $\Gamma$; representing the AGN color) given by the host-free AGN continuum, while the host will show no variation. A linear least-squares fit to the data points yields the AGN slope. The host slope passes through the origin and as it is flatter than typical AGN slopes ($\Gamma\sim1$ in adjacent bands), the intersection of the two slopes should occur in a well-defined range. This intersection range allows us to determine the host flux contribution and to calculate the host-subtracted AGN luminosity at the time of the monitoring campaign - even without the need for high spatial resolution images (\citealt{Haas2011}). In the flux-flux diagram, the host galaxy lies on the AGN slope somewhere toward its fainter end (demonstrated for a sample of 11 nearby Seyfert galaxies and QSOs by \citet{Sakata2010}; also see \citealt{Fukugita1995}). We note that in both filters, the total flux (AGN+host) contains a small contribution from the emission lines originating in the narrow-line region (NLR). Here we define the host galaxy to include the narrow emission line (NEL) contribution. \\

\begin{figure}
    \centering
    \includegraphics[width=6.18cm]{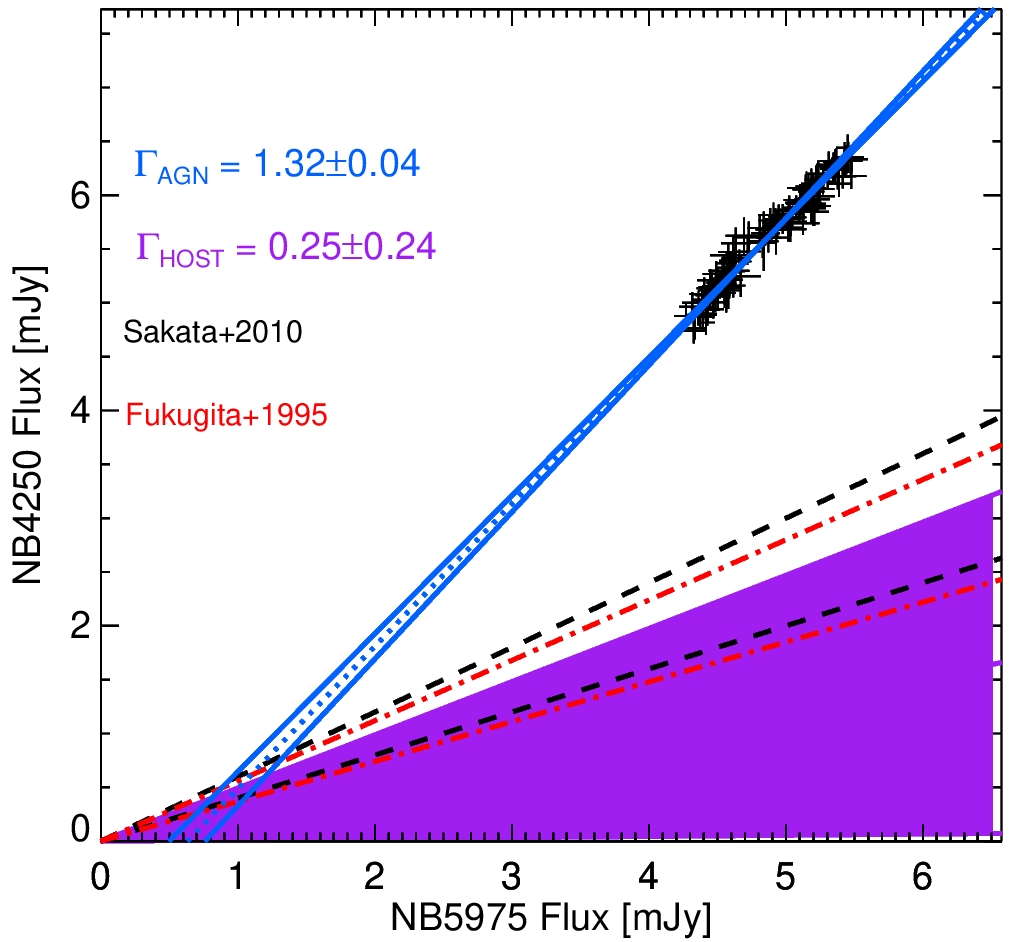}
    \includegraphics[width=6.18cm]{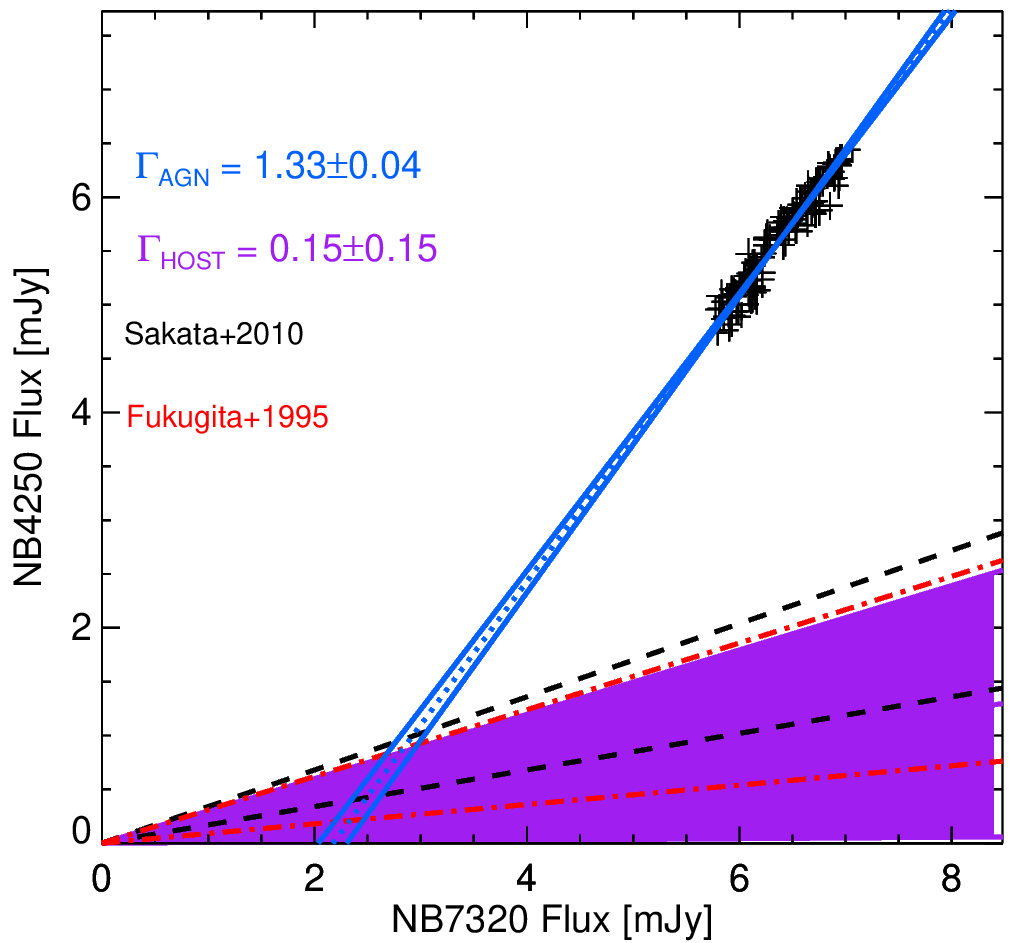}
    \includegraphics[width=6.18cm]{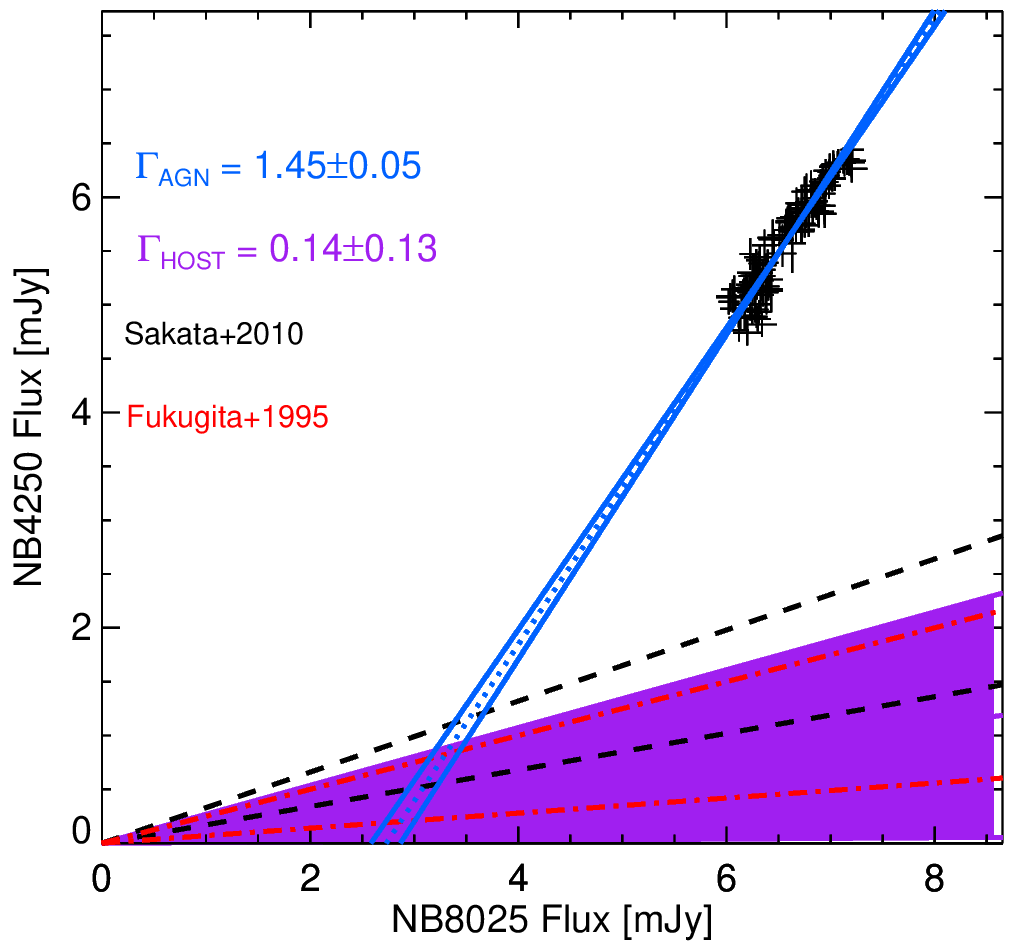}
    \caption{FVG diagram of PG 2130+099 between NB4250 and NB5975 (top), NB4250 and NB7320 (middle), and NB4250 and NB8025 (bottom). Each data point is drawn as a thin cross in which the line length corresponds to the photometric uncertainties in the respective filters. A linear least-squares fit to the data points yields the $\pm1 \sigma$ range for the AGN slope, plotted by the two steep \textit{blue} lines. The \textit{magenta} shaded area denotes the host color range from our multi-aperture photometry. The dashed lines indicate the range of the interpolated host slopes determined by \citealt{Sakata2010} (\textit{black}) and by \citealt{Fukugita1995} (\textit{red}). The intersection between the AGN and the host galaxy slope gives the host contribution in the respective band within the aperture.}
    \label{FVG}
\end{figure}

Figure \ref{FVG} shows the NB4250 versus NB5975, NB4250 versus NB7320, and NB4250 versus NB8025 fluxes of PG 2130+099. For deriving the flux-flux diagrams we used the PSF photometry light curves. Linear least-squares fits to the flux variations in each filter pair yield $\Gamma_{AGN} = 1.32\pm0.04$ for NB4250 versus NB5975, $\Gamma_{AGN} = 1.33\pm0.04$ for NB4250 versus NB7320, and $\Gamma_{AGN} = 1.45\pm0.05$ for NB4250 versus NB8025. To determine the host slope we apply multi-aperture photometry on the reference images, as proposed by \citet{Winkler1992}, where fluxes measured at different apertures
are used to estimate the host galaxy color. Since the host galaxy contribution
increases with the aperture, a linear fit between the fluxes approximates
the host slope. The total (AGN+host) fluxes for each filter are listed in Table \ref{fluxes} together with the mean host galaxy fluxes (obtained by averaging over the intersection area between the AGN and the host galaxy slopes) and the nuclear flux (calculated by subtracting the constant host galaxy component from the total flux). The uncertainties include the median errors of the calibration stars and errors caused by the black-body interpolation. The host contributes about $\sim$9\% in NB4250, $\sim$24\% in NB5975, $\sim$41\% in NB7320, and $\sim$47\% to the fluxes in NB8025. \\
\begin{table}[h]
\renewcommand{\arraystretch}{1.2}
\caption{Total (AGN+host), host galaxy, and AGN continuum fluxes.}
\begin{tabu} to 0.493\textwidth {X[c]X[c]X[c]X[c]} 

\hline
\hline 
Filter & Total (mJy) & Host (mJy) & AGN (mJy) \\
\hline 
NB4250 & $5.60\pm0.40$ & $0.50\pm0.06$ & $5.10\pm0.87$ \\ 
NB5975 & $4.86\pm0.21$ & $1.16\pm0.25$ & $3.70\pm0.96$ \\ 
NB7320 & $6.38\pm0.29$ & $2.63\pm0.23$ & $3.75\pm0.50$ \\
NB8025 & $6.58\pm0.23$ & $3.10\pm0.12$ & $3.48\pm0.26$ \\ \hline 
\end{tabu}
\label{fluxes}    
\end{table}

\begin{figure}
    \centering
    \includegraphics[width=8.3cm]{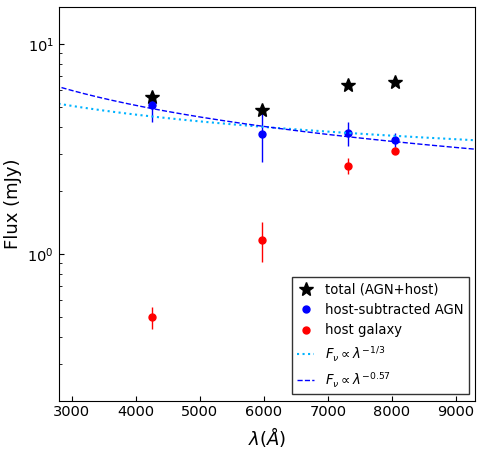}
    \caption{SED of PG 2130+099. Blue points show the host subtracted AGN continuum with a power law spectral shape of $F_\nu \propto \lambda^{-0.57\pm0.15}$ (dashed \textit{blue} line). The dotted \textit{light-blue} line corresponds to a spectral shape as predicted by a standard Shakura-Sunyaev disk (with a spectral index of $1/3$).}
    \label{SED}
\end{figure}

To derive the host-subtracted AGN flux of PG 2130+099 at rest-frame 5100\AA\ , we interpolated between the filters NB4250-NB5975 and NB4250-NB7320, adopting for the interpolation that the AGN has a power-law spectral shape ($F_\nu \propto \nu^\alpha$). At the distance of $D_L=282.6$ Mpc (\citealt{Bentz2009}), this yields a host-subtracted AGN luminosity $\lambda L_\lambda(5100\AA)=(2.40\pm0.42)\times10^{44}$ erg s$^{-1}$. The $\sim$18\% uncertainty includes the measurement errors, the uncertainty of the AGN and host slopes, and the AGN variations. The contribution of the host galaxy to the nuclear flux of PG 2130+099 has been also studied by \citet{Bentz2006,Bentz2009} through modeling the host galaxy profile (GALFIT, \citealt{Peng2002}) from the high-resolution HST images. They obtained (concerning two epochs) a monochromatic AGN luminosity of $\lambda L_\lambda(5100\AA)=(2.24\pm0.27)\times 10^{44}$ erg s$^{-1}$ and $\lambda L_\lambda(5100\AA)=(2.51\pm0.12)\times10^{44}$ erg s$^{-1}$, in good agreement with our optical luminosity estimate. Figure \ref{SED} shows the total (AGN+host) fluxes, the host-subtracted AGN fluxes, and the host fluxes as a function of wavelength. The power-law fit to the pure AGN fluxes yields $F_\nu\sim\lambda^{-\alpha}$, with $\alpha=0.57\pm0.15$, steeper than the spectral index predicted by a standard Shakura-Sunyaev disk ($\alpha=1/3$). The host-subtracted RMS spectrum (values are listed in Table \ref{var}) shows no spectral variation, which is consistent with the use of the Choloniewski diagrams. Thus, all $F_{var}$ values are consistent with each other within their uncertainties.

\subsection{Theoretical disk size}

A geometrically thin, optically thick accretion disk provides definite predictions about the theoretical lag-wavelength structure of the AGN linked to the AGN's luminosity. To check if our RM results are comparable to the lag predictions of this standard model, we use Eq. (11) of \citet{Fausnaugh2016} to estimate the expected continuum lags as a function of quasar luminosity. We note that in this work we use a simplified version of their Eq. (11) by assuming that the ratio of external to internal heating of the disk is close to zero, meaning that the contribution of the external UV/X-ray radiation of the disk is negligible compared to internal viscous dissipation in the disk. For two different light curves with wavelengths $\lambda$ and $\lambda_0$, the predicted time delay $\tau$ between the bands is given by:
\begin{equation}
    (\tau-\tau_0) = \frac{1}{c}\ \left(X\frac{k\lambda_0}{hc}\right)^{4/3} \left(\frac{3GM_{BH}\dot{M}}{8\pi\sigma}\right)^{1/3} \times\ \left[\left(\frac{\lambda}{\lambda_0}\right)^{4/3}-1\right],
\label{theod}    
\end{equation}

\noindent where $\tau_0$ is the reference time delay at $\lambda_0$, $c$ is the speed of light, $X$ is a multiplicative factor, $k$ is the Boltzmann constant, $h$ is the Planck constant, $G$ is the gravitational constant, $\sigma$ is the Stefan-Boltzmann constant, $M_{BH}$ is the mass of the central SMBH, and $\dot{M}$ is the mass accretion rate of the disk. The factor $X\sim2.5$ accounts for systematic issues in the conversion of temperature to wavelength for a given disk location since a range of radii contributes to the emission at $\lambda$. From a flux-weighted mean radius, \citet{Fausnaugh2016} derived $X=2.49$. We see that Eq. (\ref{theod}) does not tie the time delays to luminosity, but rather to $(M_{BH}\dot{M})^{1/3}$. However, for quasars whose black hole mass is not too high, and the theoretical optical spectrum of the disk may be approximated by $F_\nu \propto \nu^{1/3}$, the product $(M_{BH}\dot{M})^{2/3}$ is a direct measure of the optical luminosity. The proportionality factor between this product and the optical luminosity can be obtained from \citealt{Davis2011} (see their Eq. (7)),
\begin{equation}
     M_{BH}\dot{M} \simeq 1.4\times 10^8 \left(\frac{L_{opt}}{10^{45} \mathrm{ergs~s^{-1}}\,\cos{i}}\right)^{3/2}\, \mathrm{M_\odot^2~yr^{-1}},
\label{m}     
\end{equation}
\noindent where $L_{opt}$ is the optical luminosity. Then, plugging Eq. \ref{m} into Eq. \ref{theod}, we get for the predicted time delay (in days) relative to a reference time delay $\tau_0$ at a wavelength of $\lambda_0=5100\AA$:
\begin{equation}
(\tau-\tau_0) \simeq 2 \ \left( \frac{L_{opt}}{10^{45}\,\mathrm{ergs~s^{-1}}}\right)^{1/2}\times\ \left[\left(\frac{\lambda}{5100\,\mathrm{\AA}}\right)^{4/3}-1\right]\,\mathrm{days}.
\label{final_eq}
\end{equation}

Since the disk emission is optically thick, it depends on the (unknown) disk inclination angle, $i$. Here we adopted $\cos{i}\sim1$, in other words we assume that the accretion disk is observed nearly face-on. If we measure $\tau$ relative to a reference time delay $\tau_0$ of a light curve with effective wavelength $\lambda_0=7320$\AA, we obtain $\tau = -0.84$ days for $\lambda=4250$\AA, $\tau= -0.39$ days for $\lambda=5975$\AA, and $\tau= +0.21$ days for $\lambda=8025$\AA. Hence, the observed lags imply a disk radius that is $2-6$ times larger than the predictions from the standard thin-disk theory. Adding an external UV/X-ray term to Eq. \ref{final_eq} (as in \citealt{Fausnaugh2016}) and assuming a local ratio of external to internal heating of 1 (i.e., the X-rays and viscous heating contribute equal amounts of energy to the disk) yields $\sim18\%$ higher time lags, still unable to explain the large discrepancy between the observed and the theoretical time lags. 

\subsection{Transfer functions}
\begin{figure}[!h]
\centering
\includegraphics[width=9cm]{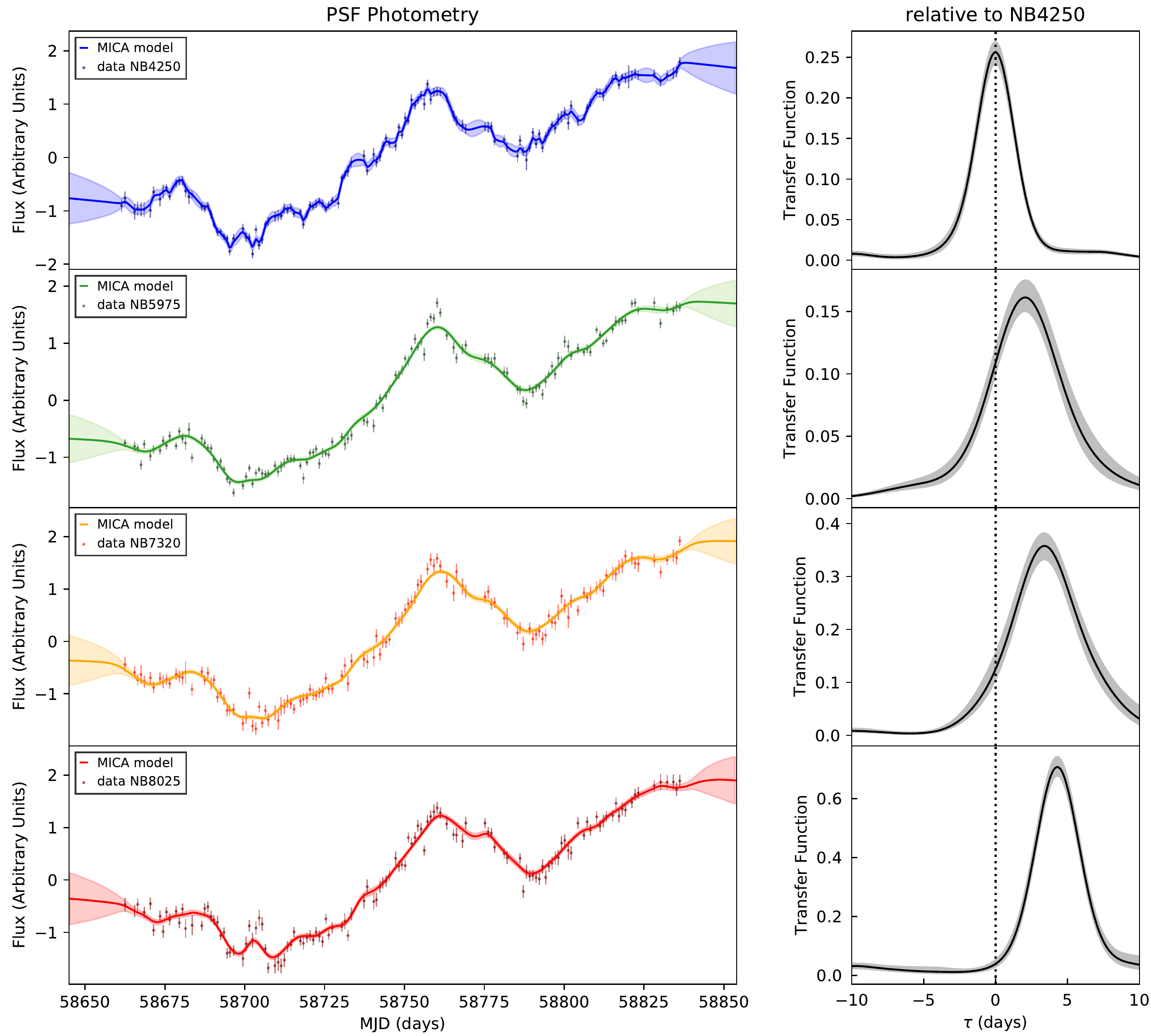}\vspace*{-2mm}
\caption{Continuum light curves at 4250, 5975, 7320, and 8025\AA\ (from top to bottom) obtained with PSF photometry (data points with error bars) overlaid with the MICA model light curves (solid lines) and uncertainties of the reconstructed light curves (shaded regions) are displayed in the left panels. The corresponding transfer functions are shown in the right hand panels with the shaded area representing the estimated uncertainties. The dotted vertical line stands for zero time delay.\vspace*{0.6cm}}
\label{mica_psf}
\centering
\includegraphics[width=9cm]{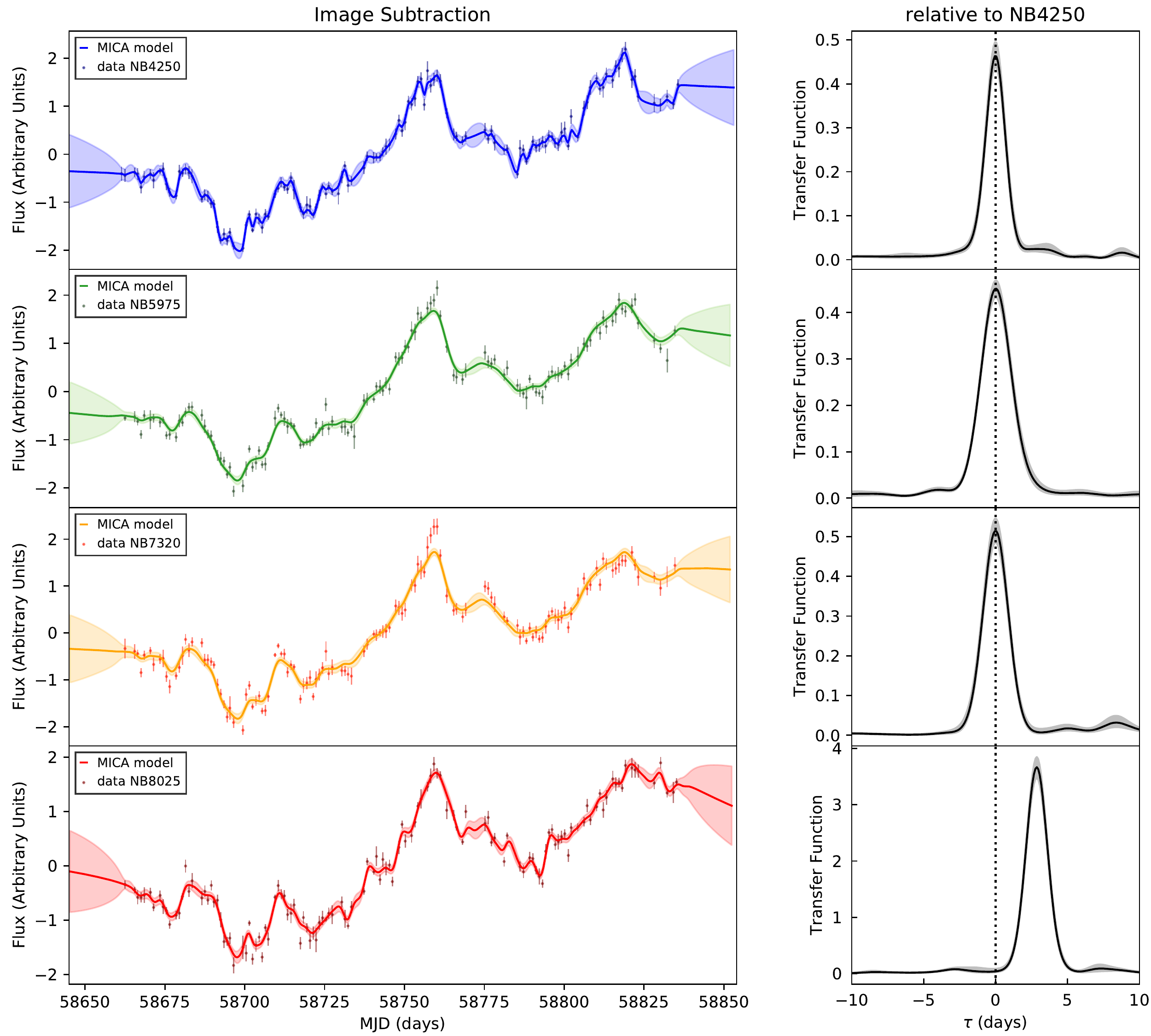}\vspace*{-2mm}
\caption{Same as Figure \ref{mica_psf} but for the continuum light curves obtained from proper image subtraction.}
\label{mica_is}
\end{figure}

The temporal behavior of spatially extended regions (outer parts of the accretion disk, BLR) are blurred echoes of the central ionizing continuum variations and directly reflect the structure of AGNs through so-called transfer functions (see \citealt{Blandford1982}). \citealt{Li2016} developed an extended, nonparametric method (MICA) to determine transfer functions for RM data based on the previous works of \citet{Rybicki1992} and \citet{Zu2011}. They express the transfer function as a sum of a collection of relatively displaced Gaussian functions, which allows arbitrary shapes of the transfer function to be accounted for (and thereby relaxes the need for presuming a specific transfer function as in previous studies, see e.g., \citealt{Rybicki1994,Zu2011}). Therefore, their approach is, to some extent, model independent and able to cope with complicated shapes of transfer functions for diverse structures and kinematics. Furthermore, the inclusion of modeling the continuum variations as a DRW process allows measurement errors to be taken into account. In RM the transfer function can be written as
\begin{equation}
    \Psi(\tau) = \sum_{k=1}^{K} f_k \exp \left[- \frac{(\tau-\tau_k)^2}{2\omega^2} \right],
\end{equation}

where $\omega$ represents the common width of Gaussian responses, $\tau_k$ and $f_k$ are the mean lag and weight of the \textit{k}th response, and $K$ is the number of responses, which can be regarded as a "smoothing parameter" (see \citealt{Li2016}). Again we set the time lag range for solving the transfer function between -10 to 10 days. We find that the best number of Gaussian functions is $K\sim8$ for the PSF photometry and $K\sim14$ for the proper image subtraction light curves. This is not surprising since the PSF photometry light curves show a smoother behavior, hence resulting in broader transfer functions. The left panels of Figure \ref{mica_psf} and \ref{mica_is} present the reconstruction of the (normalized to the mean and unit standard deviation) continuum light curves and the corresponding best recovered transfer functions measured relative to the NB4250 light curve (the bluest filter was chosen as a pivot since this band provides the best chances of resolving higher moments of the transfer function as the lag between the bluest and reddest bands is $\sim$4 times larger than the cadence) are plotted in the right panels. The proper image subtraction light curves yield narrower transfer functions peaking at smaller lags. The average of the first moment of the transfer functions (which represents the time lags) of both methods give $\tau \sim 1.1$ days for NB5975, $\tau \sim 2.0$ days for NB7320, and $\tau \sim 3.2$ days for NB8025, consistent within uncertainties with the time lags measured relative to the NB7320 filter (presented in Figure \ref{r_vs_lambda}).\\

\begin{figure}[h]
\centering
\includegraphics[width=9cm]{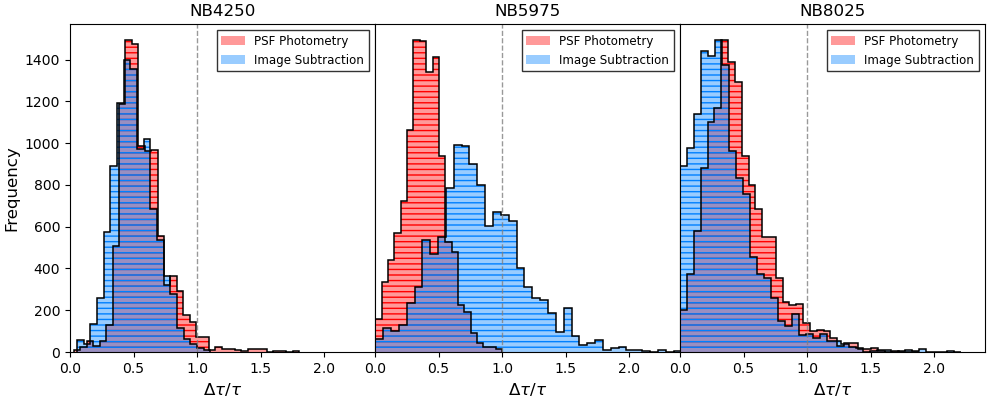}
\caption{Posterior distribution of the top-hat width obtained by JAVELIN after 10000 MCMC runs for the PSF photometry light curves (red) and proper image subtraction light curves (blue), divided by the corresponding time lag. The NB7320 light curve has been used as the reference time series. The ratio expected for a Shakura-Sunyaev accretion disk ($\Delta \tau/\tau = 1$) is indicated by the vertical dashed line.}
\label{width}
\end{figure}

Next, we used JAVELIN, which assumes a top-hat transfer function instead of Gaussian kernels, and we obtain the centroids (first moment) and widths (second moment) with respect to the NB7320 filter by means of a maximum-likelihood approach. Figure \ref{width} shows the posterior distribution of the top-hat smoothing function width $\Delta \tau$ for the PSF photometry and image subtraction light curves. The code is used in spectroscopic mode and leads to ratios between the kernel width and time delay, $\Delta \tau/\tau$, slightly smaller than unity. The observed ratios are not inconsistent with an origin of the varying continuum emission in a geometrically extended region, such as a Shakura-Sunyaev-like accretion disk.

\subsection{Directionality}
Intrinsic continuum variability is a common property in AGNs and observed in all accessible spectral bands. AGN light curves show variability over a wide range of timescales and amplitude, and many attempts have been made to determine the characteristics of these temporal flux changes. However, the origin of AGN variability may not be attributed to a single intrinsic process but a convolution of several. Thus, the physical mechanism underlying the time-dependent optical/UV variability remains poorly understood (\citealt{Giveon1999,Tachibana2020}). Among the many studies of AGN variability, several authors tried to quantify the optical variability of quasars with the hope to constrain the nature of the central engines. Some models have been proposed to explain the observed variability phenomenons, such as instabilities in the accretion disk (e.g., \citealt{Siemiginowska1997,Mineshige1994a,Mineshige1994b,Bao1996,Kawaguchi1998}), the superposition of many supernova explosions (that is to say starbust models, \citealt{Aretxaga1997}), and gravitational microlensing (\citealt{Hawkins1996, Alexander1995}). Although none of them comprehensively accounts for the observed variability and the entire range of other properties, it has been suggested that the correlation of the variability amplitude with timescale is related to intrinsic physical parameters. \\

\citet{Kelly2009} proposed a continuous time first-order autoregressive model to describe quasar optical variability. This model is also known as Ornstein-Uhlenbeck or DRW process, characterized by two parameters: the relaxation time and the variability on timescales much shorter than the relaxation time. Some studies have shown that the DRW process provides a better statistical model for most quasar variability when compared to a range of alternative stochastic/deterministic models (\citealt{Andrae2013}, but see \citealt{Mushotzky2011}). The DRW is nowadays frequently used to model aperiodic AGN light curves and a number of correlations between the DRW model parameters, the damping timescale (meaning the signal decorrelation timescale) and the variability amplitude, and the physical AGN parameters, such as luminosity and/or black hole mass, have been reported (\citealt{MacLeod2010,Kozlowski2016}). Of the two DRW model parameters, the damping timescale is of the highest interest as it can be directly linked to one of the characteristic accretion disk timescales (dynamical, thermal, or viscous;  \citealt{Czerny2006,King2008}).\\

\begin{figure}[!h]
\centering
\includegraphics[width=8.5cm]{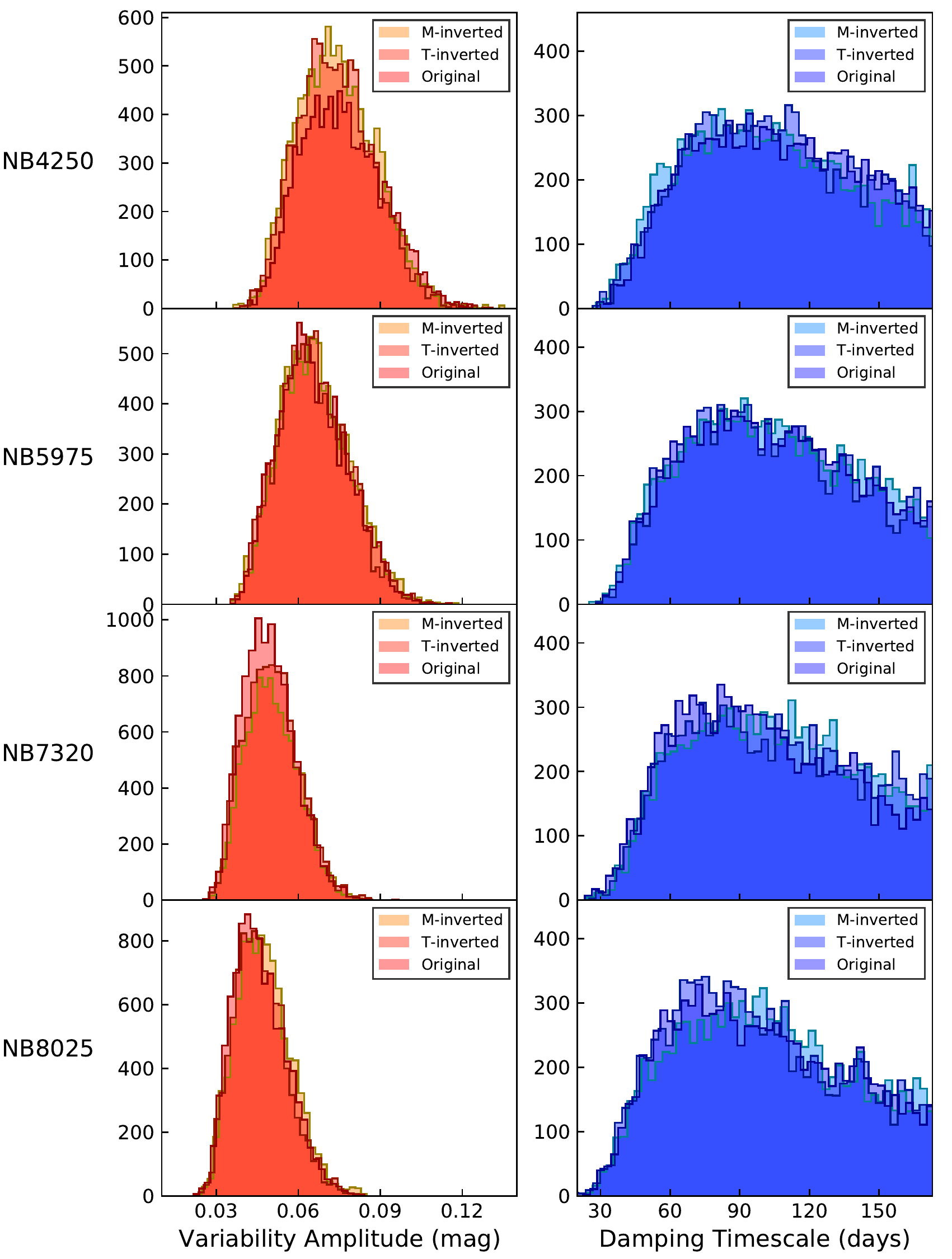}
\caption{Posterior distributions of the DRW parameters (damping timescale and modified variability amplitude) based on fits to the magnitude-inverted, time-inverted, and original continuum light curves. The apparent trend of the variability amplitude (in magnitudes) with wavelength is due to the (increasing) contribution of the host galaxy to the bands.}
\label{car}
\end{figure}

In the case of asymmetry in the continuum light curves, an autoencoder trained on quasar light curves performs differently for time-inverted (T-inverted) and magnitude-inverted (M-inverted) time series (\citealt{Tachibana2020}). We use JAVELIN to build a continuum model and to determine the DRW parameters of the original, T-inverted, and M-inverted continuum light curves. Figure \ref{car} shows the posterior distributions of the two DRW parameters for the corresponding NB light curves. The distributions for the T- and M-inverted light curves do not show any significant deviation from the distributions obtained from the original light curves, suggesting that no variability asymmetry is present. The parameter of the damping timescale is $\sim100$ days in all bands, while the variability amplitude is correlated with wavelength, showing higher amplitudes for bluer bands.\\

\begin{figure}[h]
\centering
\includegraphics[width=8.5cm]{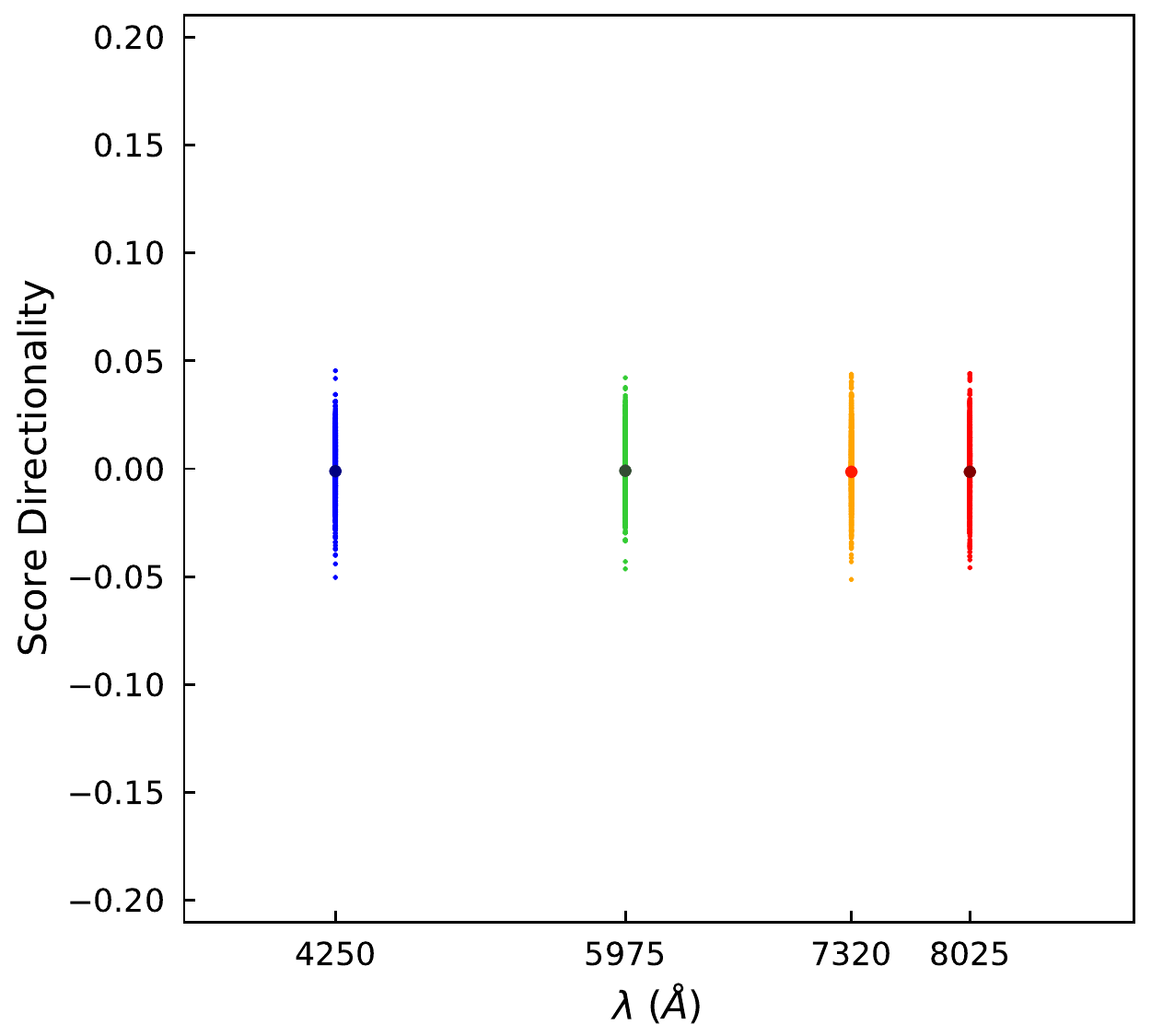}
\caption{Score of directionality (point) together with its scatter obtained via a FR-RSS formalism versus the wavelength of the corresponding continuum light curve.}
\label{score}
\end{figure}

We also estimated the directionality score (see \citealt{Lawrance1991}) for each continuum light curve, with positive values indicating that the light curve favors a rapid rise and gradual decay, and vice versa for a negative score value. Again, there is no evidence of asymmetry in the continuum light curves as shown in Figure \ref{score}. This result is consistent with the findings of \citet{Hawkins2002} who did not detect asymmetry on timescales of a year or longer in a sample of $\sim400$ quasars. However, \citet{Giveon1999} found a negative asymmetry signature (a negative asymmetry indicates that the light curve favors a rapid decay and a gradual rise) in the light curves of $\sim40$ PG quasars and more recently, \citet{Voevodkin2011} reported a significant negative asymmetry in the variations on a timescale longer than 300 days in $\sim7600$ SDSS quasars.

\section{Conclusions}\label{6}
As one of the most interesting Seyfert 1 galaxies, PG 2130+099 has been the target of several RM campaigns over the years. However, previous studies mainly focused on the delayed broad H$\beta$ line responses with respect to the varying continuum, and inconsistent time lags ranging from $\sim$10 days up to $\sim$200 days have been reported for different epochs while the optical luminosity changed no more than 40\%. To investigate the size of the continuum emitting regions, we conducted an optical RM monitoring campaign with homogeneous and high-cadence (about $\sim$1 day) observations lasting for six months from June to December 2019, using the 46cm telescope of the Wise observatory in Israel. Through a redshift of $z=0.063$, the emission line free continuum bands at 4250, 5975, 7320, and 8025\AA\ mainly trace the AGN continuum variations. Consequently, PG 2130+099 is an ideal candidate for photometric RM of the accretion disk. \\

Applying multi-aperture photometry on the reference images, we find that the spectral shape is blue (meaning the rms decreases with increasing wavelength), but is shallower than expected for a Shakura-Sunyaev accretion disk. We note that the spectrum might be reddened by AGN dust or the host's surrounding interstellar medium (ISM). However, our result ($F_\nu \propto \lambda^{-0.57\pm0.15}$) is in good agreement with the spectral slopes found for a sample of quasars (in the range of -0.6 to -0.2 in the optical-UV continuum around 3000\AA; see \citealt{Davis2007} and \citealt{Shankar2016}). We do not see any clear evidence for color changes in the quasar spectrum once the host has been subtracted off.\\

In many AGNs strong variable UV emission has been observed and the correlated variability between different bands is often discussed in the frame-work of disk reprocessing. We used the novel proper image subtraction technique to accurately measure intensity changes in a sequence of difference images. We also used traditional PSF photometry on the original images in order to compare the performance of both methods. The derived high S/N light curves are consistent with each other and we successfully detected reverberations in the NB filter light curves. We use several time-series analysis methods and report robust measurements for the continuum time lags, thereby decreasing the biases in the time-lag measurements of the accretion disk induced by single models or techniques. The time delays obtained with different methods are consistent with each other within the uncertainties and yield mean time lags of $\tau = -2.3\pm0.6$ (NB4250), $\tau =-0.9\pm0.4$ (NB5975), and $\tau =+1.3\pm0.4$ (NB8025) days with respect to the varying 7320\AA\ continuum. The time-lag measurements increase with wavelength with a power law index $\beta = 2$ but can also be fit by the relation $\tau \propto \lambda^{4/3}$ predicted for continuum reprocessing by an optically thick and geometrically thin accretion disk. However, the estimated size of the disk is a factor of $\sim$3.7 larger on average (of the three theoretical inter-band time lags) than the prediction of the standard thin-disk theory. This result challenges the current image of accretion disks and appears to corroborate those from the other recent continuum RM projects (see, e.g., \citealt{McHardy2014,Shappee2014,Lira2015,Fausnaugh2016,Jiang2017,Edelson2015,Edelson2017,Edelson2019,Cackett2018}). Most microlensing studies like \citet{Morgan2010,Mosquera2013} and \citet{Cornachione2020} found similar results suggesting that the disk sizes are significantly larger than expected. Several RM studies have been carried out using broadband filters that are contaminated by the emission of the BLR and therefore might bias the time lags to larger values. Some possible explanations for the inferred large disk sizes include disk winds leading to higher effective temperature in the outer part of the accretion disk (see for example \citealt{Sun2019,Li2019}), nonthermal disk emission cause by a low density disk atmosphere (\citealt{Hall2018}), and internal reddening of AGNs leading to an underestimation of the far-UV luminosity (\citealt{Gaskell2017}). Some authors (\citealt{Korista2001,Lawther2018}) suggest that the observed UV and optical lags do not entirely originate in the accretion disk itself but are contaminated by the diffuse continuum from the inner BLR, resulting in larger disk size measurements. Alternatively, improvements in accretion disk models could also help to resolve those discrepancies (\citealt{Dexter2011}). Unfortunately the quality of our data are not good enough to link the origin of the over-sized continuum region to a larger accretion disk, nor to provide a clear proof of a contribution of a more distant diffuse component.\\



 The intriguing result that accretion disks of AGNs might be larger than predicted by the standard thin-disk theory depends on only a handful of gravitationally lensed quasars and a few continuum RM AGNs. Given the challenges posed to standard accretion theory by these recent multi-wavelength campaigns and individual microlensing studies, it is important to carry out further continuum RM experiments and to continue studying accretion disks in a sample of AGNs covering a wide range of black hole masses, Eddington ratios and redshifts. A larger sample of AGNs with continuum lags derived to the same precision (using high-cadence NB time series) as this study would provide an interesting measurement of the lag-luminosity relation. This could provide a further test of the thin-disk theory, establish whether larger disk sizes are generic properties of the AGN population, and determine what physical parameters govern the disk size. To try to accomplish this, we are currently performing an intensive multiwavelength (in total we are using 14 NB filters) continuum RM campaign at the C18 telescope, targeting several sources on a daily basis. By increasing the number of exposures by a factor of $\sim4$ compared to the presented case, we will increase the S/N of tens of AGN light curves, hence decreasing the biases in the time-lag measurements and providing better chances of resolving higher moments of the transfer function to study the origin and kinematics of the AGN continuum emitting regions.\\

\begin{acknowledgements}
This work was financially supported by the DFG grant HA3555-14/1 to Tel Aviv University and University of Haifa. Tiffany Lewis's contribution to this work was supported by the Zuckerman STEM Leadership Program. This research has been partly supported by the Israeli Science Foundation grant no. 2398/19. 
\end{acknowledgements}
\bibliographystyle{aa}
\bibliography{bibliography}

\end{document}